%% file: main.tex
\documentclass[compsoc,conference,a4paper,10pt,times]{IEEEtran}
\IEEEoverridecommandlockouts
\usepackage{cite}
\usepackage{amsmath,amssymb,amsfonts}
\usepackage{algorithmic}
\usepackage{graphicx}
\usepackage{textcomp}
\usepackage{xcolor}
\usepackage[colorlinks=true,urlcolor=black]{hyperref}
\usepackage{multirow}
\usepackage{subcaption}
\usepackage{caption}
\usepackage{booktabs}
\usepackage{wasysym}
\usepackage{url}
\usepackage{float}
\usepackage{makecell}
\usepackage{afterpage}
\usepackage{changepage}
\usepackage{listings}
\usepackage{bmpsize}
\usepackage{tcolorbox}
\usepackage{enumitem}
\definecolor{setupgray}{rgb}{0.95, 0.95, 0.95}
\newtcolorbox{setupbox}[1][]{
    colback=setupgray,
    colframe=setupgray,
    arc=3pt,
    boxrule=0pt,
    left=6pt, right=6pt, top=6pt, bottom=6pt,
    fonttitle=\bfseries,
    #1
}

\def\BibTeX{{\rm B\kern-.05em{\sc i\kern-.025em b}\kern-.08em
    T\kern-.1667em\lower.7ex\hbox{E}\kern-.125emX}}

\begin{document}
\title{SoK: Security Evaluation of Wi-Fi CSI Biometrics: Attacks, Metrics, and Open Challenges}

\author{
    \IEEEauthorblockN{
        Gioliano de Oliveira Braga, Pedro Henrique dos Santos Rocha, Rafael Pimenta de Mattos Paixão,\\
        Giovani Hoff da Costa, Gustavo Cavalcanti Morais, Lourenço Alves Pereira Jr.
    }
    \IEEEauthorblockA{Division of Computer Science, Aeronautics Institute of Technology (ITA), Brazil\\
    \{giolianobraga,ljr\}@ita.br}
}

\maketitle

\begin{abstract} Wi-Fi Channel State Information (CSI) has been repeatedly proposed as a biometric modality, often with reports of high accuracy and operational feasibility. However, the field lacks a consolidated understanding of its security properties, adversarial resilience, and methodological consistency. This Systematization of Knowledge (SoK) examines CSI-based biometric authentication through a security lens, analyzing how existing works diverge in sensing infrastructure, signal representations, feature pipelines, learning models, and evaluation methodologies. Our synthesis reveals systemic inconsistencies: reliance on aggregate accuracy metrics, limited reporting of FAR/FRR/EER, absence of per-user risk analysis, and scarce consideration of threat models or adversarial feasibility.

To this end, we construct a unified evaluation framework to expose these issues empirically and demonstrate how security-relevant metrics such as per-class EER, Frequency Count of Scores (FCS), and the Gini Coefficient uncover risk concentration that remains hidden under traditional reporting practices. The resulting analysis highlights concrete attack surfaces—including replay, geometric mimicry, and environmental perturbation—and shows how methodological choices materially influence vulnerability profiles. Based on these findings, we articulate the security boundaries of current CSI biometrics and provide guidelines for rigorous evaluation, reproducible experimentation, and future research directions. This SoK offers the security community a structured, evidence-driven reassessment of Wi-Fi CSI biometrics and their suitability as an authentication primitive.
\end{abstract}


\section{Introduction}
\label{sec:intro}

User authentication remains one of the core challenges in modern cybersecurity \cite{Liu2014}. Physical-layer biometrics, which rely on intrinsic physiological and behavioral traits, offer a compelling alternative by removing the need for explicit user actions or dedicated sensing hardware.

Wi-Fi Channel State Information (CSI) has gained prominence as a versatile sensing modality for human-centric applications, capturing subtle variations in wireless propagation caused by a person’s presence, movements, and intrinsic attributes \cite{Shah2017, Ma2019}. In contrast to conventional RSSI, CSI exposes amplitude and phase measurements at the subcarrier level, providing much finer spatial and temporal resolution. Its complex-valued structure encodes distinctive electromagnetic patterns shaped by an individual’s body geometry, tissue properties, and attitude patterns.

The literature on CSI-based biometric authentication has evolved along multiple dimensions. Dynamic authentication systems exploit temporal behavioral patterns such as gait \cite{Shi2017, Shi2021}, keystroke dynamics \cite{Gu2022}, and gestures \cite{Zhang2022}. These approaches leverage spatiotemporal CSI variations to capture unique user behaviors, demonstrating high accuracy \cite{Gu2022}. However, dynamic systems inherently face challenges related to motion artifacts and the need for consistent behavioral patterns. In contrast, static authentication systems focus on geometric and radiometric properties captured through stable electromagnetic interactions, such as palm placement \cite{Liu2014, Shah2017, trindade2025}. These approaches help reduce motion artifacts and yield more repeatable measurements, making them suitable for scenarios that require consistent positioning—identifying a person by who they are rather than by what they know or do. Behavioral authentication systems extend this by exploiting walking and stationary patterns during daily activities for continuous authentication, though they face challenges from environmental interference and multiple individuals \cite{Gu2021}. Multi-frequency approaches have been explored to enhance robustness and discriminative power by combining information from multiple frequency bands \cite{Han2023}, albeit requiring more complex hardware and increased computational complexity. Advanced applications of CSI sensing also extend beyond authentication to include 3D human pose estimation \cite{Geng2022, Ramesh2025} and secret key generation \cite{Srinivasan2022}. These advanced approaches, however, rely on sophisticated deep learning architectures and substantial computational resources.

Unfortunately, despite substantial progress in CSI-based research, several critical challenges persist in the broader field of physiological biometric authentication. The area still suffers from a limited number of studies, fragmented methodologies, and inconsistent validation practices, making it difficult to assess whether reported improvements reflect real and meaningful advances. Reproducibility issues are common, often stemming from dependence on proprietary hardware (e.g., the Intel 5300 NIC) or insufficient implementation details. Moreover, many studies report only basic accuracy metrics while overlooking essential security-oriented measures such as Equal Error Rate (EER), False Acceptance Rate (FAR), and False Rejection Rate (FRR). These omissions frequently occur alongside a lack of auxiliary statistical analyses—such as score-distribution modeling or subject-level fairness evaluation—which are crucial for ensuring robustness in real-world deployments.

This Systematization of Knowledge (SoK) addresses these gaps by introducing a unified framework for Wi-Fi CSI–based static physiological biometric authentication, emphasizing reproducibility and standardization. We systematize fragmented methodologies across the literature, revealing how methodological choices influence security properties. Our approach combines physics-based principles—rooted in Maxwell’s equations and their impact on amplitude, phase, and energy profiles—with data-driven techniques such as statistical filtering, normalization, and security-focused evaluation metrics.

\subsection{Motivation}
\label{subsec:motivation}

Wi-Fi CSI offers a non-intrusive, cost-effective biometric modality that leverages existing infrastructure, avoiding the dependencies of specialized hardware~\cite{Vieira2025}. Extractable via open-source tools (e.g., Nexmon~\cite{nexmon_repo}) on commodity devices, it provides a scalable complementary layer for continuous authentication where explicit user interaction is impractical.

However, transitioning from experimental feasibility to secure production demands rigorous adversarial resilience. Unlike cryptography, wireless biometric entropy is stochastic and vulnerable to drift, replay, and data reconstruction attacks~\cite{Wen2025}. Relying solely on aggregate metrics creates a ``false sense of security,'' masking critical vulnerabilities in specific user subsets. Validating CSI as a trustworthy primitive requires transcending mean accuracy to evaluate distributional robustness and consistency. This necessity, coupled with the IEEE 802.11bf standardization, drives the need for a framework that addresses these structural gaps.

\subsection{Scope and Objective}
\label{subsec:scope_obj}

To provide a reproducible framework for static physiological biometric authentication, this work focuses on the physical-layer representation of the wireless channel as a means of identifying individuals through their electromagnetic interaction with the propagation field. Our study targets \textbf{static palm-based identification for robust authentication}, with the subject’s hand positioned approximately 5~cm above the Raspberry~Pi~4B antenna—capturing fine radiometric and geometric patterns while minimizing motion artifacts and ensuring repeatable measurements.

The experiment used the 5~GHz Wi-Fi band, where the increased granularity and field perturbations enhance the discriminative power of CSI amplitude and phase features~\cite{Guan2023}. In addition, this band exhibits lower amplitude variability—and thus reduced noise—compared to 2.4~GHz~\cite{Zhuravchak2022}, while offering superior stability and better temporal resolution under a 40~MHz bandwidth than under 20~MHz~\cite{Tian2018, Kong2021}. Together, these factors contribute to improved model performance~\cite{Meneghello2025}.

Another characteristic of lower Wi-Fi frequencies is their relationship with Free-Space Path Loss (FSPL). Although these bands provide longer propagation range, they are also more susceptible to environmental interference due to the widespread presence of consumer electronic devices in everyday settings~\cite{Tian2018, Ma2019}.

The primary objective is to unify the fragmented methodologies found in the literature into a coherent pipeline that spans raw CSI acquisition, feature extraction, and machine learning classification. In doing so, the approach bridges the gap between theoretical electromagnetics and practical biometric intelligence.

Within this scope, the study aims to establish a consistent experimental foundation using commodity Wi-Fi hardware and open-source tools (notably Nexmon). It also seeks to define a unified preprocessing and feature-extraction methodology that preserves physical interpretability. The research then evaluates multiple classifiers under controlled conditions to assess their discriminative power and robustness.

By framing CSI as both a physical quantity and a biometric “meter”, this work advances the systematization of CSI-based static physiological authentication research, providing a transparent, replicable, and physics-aware benchmark for future studies.

\subsection{Contributions}
\label{subsec:contributions}

In summary, this work makes the following key contributions:

\begin{description}[style=unboxed, leftmargin=0cm, font=\bfseries]

    \item[Systematization and Benchmarking:] 
    We propose a five-dimensional taxonomy and perform a comparative benchmark of state-of-the-art (SOTA) systems.

    \item[Reproducible Methodology:] 
    We systematize evaluation methodologies, establishing a framework for reproducible and security-oriented assessment.

    \item[Security-Oriented Framework:] 
    We define an evaluation framework prioritizing security metrics beyond accuracy, specifically introducing per-class EER, FCS, and the Gini Coefficient (GC).

    \item[Vulnerability and Risk Analysis:] 
    We conduct a systematic security analysis exposing critical threats:
    \begin{itemize}[leftmargin=1em, nosep, label=\textbullet]
        \item \textbf{Attack Surface:} Identification of replay attacks, geometric mimicry, and environmental threats.
        \item \textbf{Per-User Risk:} The first analysis demonstrating uniform security requirements and exposing specific per-user vulnerability patterns.
    \end{itemize}

\end{description}

The remainder of this paper is organized as follows. Section~\ref{sec:background} introduces the fundamentals of Wi-Fi CSI and the theoretical basis for feature extraction. Section~\ref{sec:taxonomy} presents a structured taxonomy of CSI-based biometric systems. Section~\ref{sec:related} provides a comparative benchmark of prior work and identifies critical gaps. Section~\ref{sec:methodology} describes the unified, reproducible pipeline. Section~\ref{sec:results} reports the experimental validation of static palm-based authentication. Section~\ref{sec:discussion} examines key strengths, limitations, and security implications. Section~\ref{sec:future} outlines future directions, including anomaly detection and the impact of IEEE~802.11bf on Wi-Fi Sensing (WFS). Finally, Section~\ref{sec:conclusion} summarizes the contributions and positions this SoK as a reference for advancing practical CSI-based biometric systems.

\section{Background on Wi-Fi CSI and Biometric Sensing}
\label{sec:background}

This section introduces the fundamental principles of Wi-Fi CSI and the theoretical basis for feature extraction in biometric authentication. We present the core signal models, phase calibration procedures, and key feature categories derived from amplitude, phase, and energy domains. The remaining, more complex, hand-crafted feature formulations are detailed in Appendix~\ref{app:feature-extraction}.

In real-world environments, wireless signals propagate along multiple paths due to reflections and refractions. The received spectrum, $R(f)$, is related to the transmitted spectrum, $S(f)$, through the frequency-domain relationship~\cite{Shah2017, Tang2020, Hernandez2022}:
\begin{equation}
    R(f) = S(f) H(f) + N(f)
    \label{eq:raw_CSI}
\end{equation}
where $H(f)$ represents the CSI. This equation serves as the basis for extracting complex CSI data using the Nexmon tool (Section~\ref{subsec:aqc-tool}).

The Channel Frequency Response (CFR), $H(f)$, represents the complex transfer function of the channel~\cite{Ma2019, Tang2020, Hernandez2023}:
\begin{equation}
    H(f) = |H(f)| e^{j\phi(f)}
\end{equation}
where $|H(f)|$ denotes the magnitude response and $\phi(f)$ the phase response. Its time-domain equivalent, the Channel Impulse Response (CIR), is given by~\cite{OYERINDE2018, Song2018, tsinghua_tutorial}:
\begin{equation}
    h(t) = \sum_{n=1}^{N} \alpha_n e^{-j\phi_n} \delta(t - \tau_n)
\end{equation}
where $\alpha_n$, $\phi_n$, and $\tau_n$ denote the complex attenuation, phase, and time delay of the $n^{\text{th}}$ path.

\subsection{Phase Calibration Formulas}

Phase calibration systematically removes hardware artifacts through sequential stages to restore the intrinsic phase response encoding user-specific geometric interactions~\cite{Hu2023, Ma2019, Kong2023}.

\textbf{Non-linear Phase Calibration:} Following the Tsinghua University methodology~\cite{tsinghua_tutorial}, this multi-stage process compensates for hardware-induced distortions. It corrects errors like Packet Detection Delay (PDD) that introduce unpredictable drifts, obscuring genuine channel characteristics.

\textbf{CFO Removal:} Carrier Frequency Offset (CFO) arises from oscillator mismatches, causing a constant phase shift. To mitigate this, a global offset is subtracted:
\begin{equation}
    H_{\text{calib}}(f) = H(f) \cdot e^{-j\phi_{\text{offset}}}
\end{equation}
where $\phi_{\text{offset}} = \text{median}(\phi(f))$ is the median phase bias. This re-establishes a consistent spectral reference reflecting propagation components rather than hardware drift.

\textbf{Phase Unwrapping:} Since CSI phase is confined to $[-\pi, \pi]$, abrupt $2\pi$ discontinuities can occur. Unwrapping reconstructs a continuous profile:
\begin{equation}
    \phi_{\text{unwrapped}}(f_k) = \text{unwrap}(\phi(f_k))
\end{equation}
This step preserves physical variations caused by path-length differences or fine-grained movements that would otherwise be masked by artificial wraps.

\textbf{Detrending:} To remove linear slopes induced by SFO and PDD, a linear regression trend is estimated and subtracted:
\begin{equation}
    \phi_{\text{detrended}}(f_k) = \phi_{\text{unwrapped}}(f_k) - \text{linear\_trend}(f_k)
\end{equation}
This isolates residual fluctuations associated with the propagation environment, focusing analysis on geometry-dependent scattering.

\textbf{Normalization:} Finally, the phase is mean-centered to remove global offsets:
\begin{equation}
    \phi_{\text{normalized}}(f_k) = \phi_{\text{detrended}}(f_k) - \mathbb{E}[\phi_{\text{detrended}}(f_k)]
\end{equation}
The resulting zero-mean distribution emphasizes relative phase characteristics, capturing micro-scale variations induced by subject geometry—key discriminative signatures in CSI biometrics~\cite{Shi2021, yang2022handson, Avola2022}.

\subsection{Magnitude-Based Features: Ampère-Maxwell Law}
\label{subsec:magnitude}

Measured amplitude variations reflect electromagnetic energy fluctuations induced by the subject's body geometry~\cite{Geng2022}. Analyzing these features quantifies the dynamic redistribution of field intensity, serving as a proxy for radiometric signatures~\cite{Avola2022}.

\textbf{Fundamental Magnitude Calculation:} Extracted from complex CSI, the amplitude response is~\cite{Zhou2019, Almeida2025, trindade2025}:
\begin{equation}
    |H(f)| = \sqrt{\text{Re}^2\{H(f)\} + \text{Im}^2\{H(f)\}}
\end{equation}

\textbf{Amplitude Mean Features:} The mean amplitude across subcarriers and sample blocks is computed as shown in (\ref{eq:amp-mean}).
\begin{equation}
    \text{amp\_mean} = \frac{1}{K} \sum_{k=1}^{K} \frac{1}{T} \sum_{t=1}^{T} |H(f_k, t)|
    \label{eq:amp-mean}
\end{equation}
where $K$ and $T$ denote the number of subcarriers and time samples, respectively. Other features like skewness (\ref{eq:skew_amp}) and kurtosis (\ref{eq:kur_amp}) characterize the palm’s reflection pattern, capturing spectral asymmetry caused by physiological factors (tissue thickness, bone structure). This forms a unique “palmprint” frequency signature~\cite{Chowdhury2017, Shi2021, Bisio2024}.

\subsection{Phase-Based Features: Faraday Law}
\label{subsec:phase}

Phase components were processed using the heuristics detailed in Section~\ref{sec:methodology} to derive corresponding features.

\textbf{Fundamental Phase Calculation:} The phase response is defined as~\cite{Almeida2025}:
\begin{equation}
    \phi(f) = \arctan\left(\frac{\text{Im}\{H(f)\}}{\text{Re}\{H(f)\}}\right)
\end{equation}

\textbf{Phase Mean Features:} The mean phase across subcarriers is defined as:
\begin{equation}
    \text{phase\_mean\_mean} = \frac{1}{K} \sum_{k=1}^{K} \frac{1}{T} \sum_{t=1}^{T} \phi(f_k, t)
\end{equation}

The mean and standard deviation of phase standard deviations are computed as follows~\cite{Lv2018, Cheng2022, Lin2023}:

\begin{equation}
\begin{split}
    \text{phase\_std\_mean} = & \\
    \frac{1}{K} \sum_{k=1}^{K} \sqrt{\frac{1}{T-1} \sum_{t=1}^{T} \left(\phi(f_k, t) - \frac{1}{T} \sum_{t'=1}^{T} \phi(f_k, t')\right)^2}
\end{split}
\end{equation}

\begin{equation}
\begin{split}
    \text{phase\_std\_std} = & \\
    \sqrt{\frac{1}{K-1} \sum_{k=1}^{K} \left(\text{Std}_k\{\phi(f_k, t)\} - \text{phase\_std\_mean}\right)^2}
\end{split}
\end{equation}

These features leverage the phase gradient (\ref{eq:phase-grad}) to mitigate errors like CFO and SFO~\cite{Li2024.1}. Dispersion analysis (expressions~\ref{eq:dphi_std_mean} and~\ref{eq:dphi_std_std}) quantifies unique phase textures shaped by user geometry, enabling reliable authentication even in dynamic environments~\cite{Liu2019, Cheng2021, Wei2025}. Establishing a robust phase methodology is crucial for extracting stable environmental cues and improving classification performance~\cite{Diaz2023}.

\subsection{Energy-Based Features: Poynting Theorem}

This methodology extracts features based on average reflected and absorbed power, capturing individual-specific scattering patterns and quantifying hand-induced scattering selectivity~\cite{Li2022, Pu2023}.

\textbf{Fundamental Energy Calculation:} The energy per subcarrier is computed from the squared magnitude~\cite{Dai2020, Xu2022}:
\begin{equation}
    E(f_k) = \frac{1}{T} \sum_{t=1}^{T} |H(f_k, t)|^2
\end{equation}
$E(f_k)$ estimates energy propagated, reflected, and absorbed by the body solely from CSI observations. Skewness (\ref{eq:energy-skewness}) and kurtosis (\ref{eq:energy-kurtosis}) metrics characterize energy distribution by capturing body-contour effects, while energy entropy (\ref{eq:energy-entropy}) quantifies uniformity, providing a robust authentication feature~\cite{Li2020, Guan2023}.

\textbf{Spectral Features (Wave Equation):} Spectral features (\ref{eq:spec-centroid}--\ref{eq:spec-width}) quantify CSI energy distribution across subcarriers. Essential for capturing fine-grained patterns and distinguishing spectral noise (\ref{eq:spec-flatness}), they support high-resolution recognition in Wi-Fi CSI systems~\cite{Li2020.1, Nakamura2020, Tan2022, Chen2023, Ratman2024}.

\subsection{Empirical Energy Features: Energy Conservation}

These features decompose CSI energy into three components: Reflection (\ref{eq:reflect-energy}), Absorption (\ref{eq:absorv-energy})—linked to biophysical factors—and Refraction (\ref{eq:refract-energy}), associated with phase variation. Normalizing this decomposition (\ref{eq:emp-norm-refl}--\ref{eq:emp-norm-refr}) yields a unique, forgery-resistant physical signature critical for robust fine-grained recognition~\cite{Zhang2019, Wu2021, Avola2022, Wu2023, Li2024, Liu2014}.

\subsection{Temporal Variability Features}

Temporal Variability (\ref{eq:temp-variat-mean}--\ref{eq:temp-variat-cv}) and Stability (\ref{eq:stab-measure}--\ref{eq:coeff-variat-std}) metrics quantify CSI amplitude fluctuations over time. Capturing characteristic physiological and behavioral variations, these features facilitate activity segmentation and robust user profiling~\cite{Oshiga2019, Meng2023, Wang2023}. Despite our focus on static authentication, these metrics effectively distinguish subtle inter-user differences, enhancing precision and specificity.

\subsection{Stability \& Spatial Correlation Features}

\textbf{Stability Features:} These quantify temporal consistency of CSI amplitude across subcarriers. By measuring magnitude uniformity via the coefficient of variation, they enable discrimination based on unique temporal coherence signatures.

\textbf{Spatial Correlation Features:} Evaluating coherence across adjacent subcarriers (\ref{eq:adj-subc-corr}--\ref{eq:adj-subc-std}), this metric filters noisy subcarriers to accurately represent physical interactions (radiobiometrics), enhancing robustness against fraud~\cite{Shi2017, Ding2019, Meneghello2022, Lu2023}.

\textbf{Spectral Roughness Analysis:} Defined in (\ref{eq:roughness}--\ref{eq:spec-rough-std}), roughness quantifies fine-grained spectrum irregularities from multipath interactions and morphological differences~\cite{Satija2019, Chen2023}. The $\text{spectral\_roughness\_std}$ (\ref{eq:spec-rough-std}) indicates scattering diversity: high values denote irregular transitions, while lower values reflect uniformity, offering stable user-specific signatures~\cite{Sharma2024}.

\textbf{Spectral Curvature:} Complementing roughness, spectral curvature (\ref{eq:spec-curve}--\ref{eq:spec-curve-std}) measures the rate of change in local variations, characterizing the spectral envelope's ``bending''~\cite{Wang2024}. Distinct body shapes induce unique curvature patterns, serving as high-resolution frequency-domain signatures~\cite{Wu2023}.

\begin{figure*}[t!]
    \centering
    \includegraphics[width=0.9\linewidth, height=6cm, keepaspectratio]{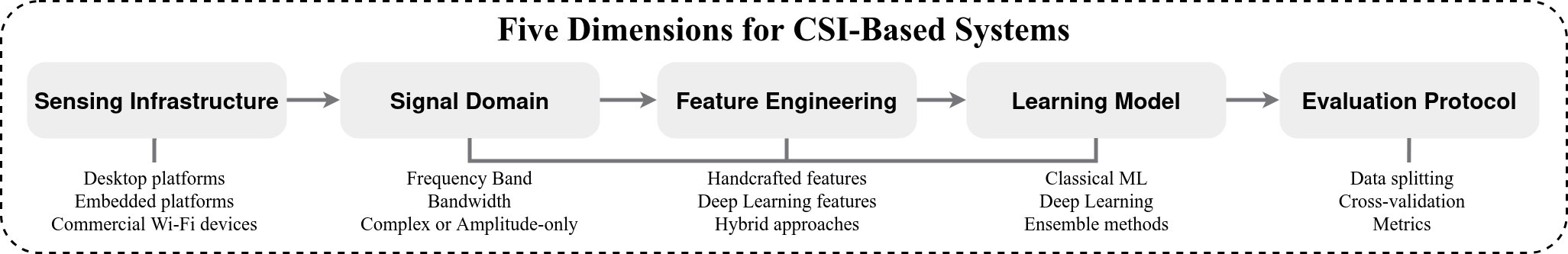}

    \caption{A Taxonomy of CSI-Based biometric systems.}
    \label{fig:taxonomy}
\end{figure*}

\subsection{CSI Acquisition Toolchain \& Experimentation Setup}
\label{subsec:aqc-tool}

CSI acquisition utilized a customized open-source toolchain based on the Nexmon framework~\cite{trindade2025, Guan2023, Turetta2022}, enabling CSI extraction on embedded IoT systems. Unlike legacy desktop platforms (Intel~5300~\cite{Gu2021}, Atheros~\cite{Fukushima2022}), this approach provides a fully portable, low-cost sensing platform on the \textbf{Raspberry~Pi~4B}.

\textbf{Hardware Components:} The experimental setup comprised three primary devices:
\begin{setupbox}
    \begin{description}[style=unboxed, leftmargin=0cm, parsep=3pt, font=\bfseries]
        \item[Access Point (Transmitter):] TP-Link Archer C60 configured as AP in \textbf{IEEE~802.11n} mode on \textbf{Channel~36} (5180\,MHz) with \textbf{40\,MHz} bandwidth.
        \item[CSI Receiver (Sniffer):] Raspberry~Pi~4B (\texttt{BCM43455c0} chipset) in \textbf{monitor mode} (1\,dBm Tx power). Captured raw CSI from 128 OFDM subcarriers at 5180\,MHz via a custom script.
        \item[Client (Traffic Generator):] Samsung Galaxy Tab~S9~FE (\textbf{64-QAM}) running \textbf{iPerf2}, generating UDP traffic to induce CSI variations.
    \end{description}
    \vspace{0.3em}
    \noindent\small\textit{Note:} The receiver was powered by a \textbf{Peining WUP-945} power bank (30,000\,mAh, 22.5\,W) via 5\,V/3.0\,A USB output.
\end{setupbox}

\textbf{Software and Configuration:} The receiver utilized the \textbf{Nexmon~CSI} toolchain~\cite{nexmon_repo, Grigoli2019} for Raspberry~Pi kernel~5.10.92. Firmware hooks extracted the raw complex-valued CSI matrix $H(f,t)$ for each subcarrier. The iPerf2 client transmitted a continuous UDP flow using:
\begin{equation}
\texttt{\shortstack[l]{
iperf -c 192.168.1.1 -u -b 500M -t 60\\
\hspace{3.6em}-i 1 -l 1400 -p 5500
}}
\label{eq:iPerf2}
\end{equation}
This configuration (500\,Mbit/s, 60s, 1400-byte datagrams) ensured channel saturation. Leveraging the dataset from~\cite{trindade2025}, each subject performed a two-minute acquisition (five samples per hand), maintaining a static hand position \textbf{5\,cm above the antenna}.

\textbf{Operational Overview:} The tablet transmitted UDP packets to the router while the Raspberry~Pi passively captured CSI from Wi-Fi frames. Data was transferred via Ethernet/SSH to an MSI GS63 laptop~\cite{msi_gs63_8re} for storage. CSI data, stored as \texttt{.pcap} files, were parsed into complex matrices $H(f_k,t)$ (128 subcarriers) via Nexmon’s Python interface.

\section{Taxonomy of CSI-Based Biometrics}
\label{sec:taxonomy}

This section systematizes prior research on CSI-based biometrics through a five-dimensional taxonomy: sensing infrastructure, signal domain, feature engineering, learning model, and evaluation protocol (see Figure~\ref{fig:taxonomy}. We review state-of-the-art approaches to establish a coherent understanding of the field's evolution. Within this framework, our work—utilizing a Raspberry~Pi~4B at 5\,GHz with a \textbf{static acquisition design}—is categorized as a device-based, closed-set identification system for authentication relying on handcrafted features and standard classifiers (RandomForest/MLP).

\subsection{Scope and Domain}
\label{subsubsec:scope-domain}

CSI-based biometric systems generally fall into three primary domains, categorized by their spatial and temporal dependencies.

\textbf{Static Authentication} focuses on identifying users based on stationary body placement, such as hand posture or sitting position. These systems emphasize geometric and radiometric signal properties rather than motion artifacts~\cite{Liu2014, Shah2017}.

In contrast, \textbf{Dynamic Authentication} leverages temporal behavioral patterns—including gait, gestures, or breathing activity—by exploiting spatiotemporal CSI variations over time~\cite{Shi2017, Shi2021, Gu2022}.

Finally, \textbf{Multi-User Authentication} addresses the complex challenge of distinguishing multiple concurrent users within a shared channel. This domain often necessitates advanced signal separation techniques to isolate individual biometric signatures~\cite{Kong2023, Kong2021}.

The choice of domain fundamentally dictates system design: static systems prioritize spatial resolution and feature stability, whereas dynamic systems require robust sequence modeling and motion sensitivity.

\subsection{Task Definition}
\label{subsubsec:task-def}

CSI-based biometrics primarily address three tasks.

\textbf{Authentication (Verification)} involves one-to-one matching, relying on FAR/FRR as key metrics~\cite{Kong2022, Han2023, Lin2023}.

\textbf{Identification} performs one-to-many classification, typically evaluated via precision, recall, ROC-AUC, and per-class EER using a One-vs-Rest strategy~\cite{Guan2023, Kong2023}.

\textbf{Continuous Monitoring} enables long-term authentication, requiring drift handling, adaptive thresholds, or session tracking~\cite{Wu2021, Cheng2022}.

While authentication emphasizes security (low FAR), identification targets discriminability across users. Our work focuses on closed-set identification for authentication using static palm-based biometrics.

\input{benchmark_table}

\subsection{Classification Dimensions}
\label{subsubsec:class-domain}

Our taxonomy organizes CSI-based biometric systems along five core dimensions, detailed below.

\begin{description}[style=unboxed, leftmargin=0cm, font=\bfseries]

    \item[1\textsuperscript{st} Sensing Infrastructure:] 
    This dimension outlines the hardware toolchain for CSI extraction:
    \begin{itemize}[leftmargin=1em, nosep, label=\textbullet]
        \item \textbf{Desktop platforms:} Utilize Intel~5300 or Atheros NICs requiring desktop-class hardware~\cite{Gu2021, Gu2021.1, Lu2022, Fukushima2022}.
        \item \textbf{Embedded platforms:} Solutions like Raspberry~Pi (Nexmon)~\cite{trindade2025, Guan2023, Turetta2022} or ESP-based toolkits~\cite{Basha2024}, enabling portable IoT deployments.
        \item \textbf{Commercial Wi-Fi devices:} Off-the-shelf routers trading fine-grained CSI access for deployment ease~\cite{Meng2023}.
    \end{itemize}

    \vspace{0.4em}
    \item[2\textsuperscript{nd} Signal Domain:] 
    Systems differ in frequency, bandwidth, and representation:
    \begin{itemize}[leftmargin=1em, nosep, label=\textbullet]
        \item \textbf{Frequency band:} 5\,GHz offers finer spatial resolution compared to 2.4\,GHz due to wavelength interactions~\cite{Guan2023, Kong2021, Gu2021}.
        \item \textbf{Bandwidth:} Wider bands (20/40/80\,MHz) increase subcarrier density and spectral richness~\cite{Tian2018, Kong2021, Meneghello2025}.
        \item \textbf{Representation:} Choices between amplitude-only, phase-only, or complex CSI determine the captured physical phenomena~\cite{Lin2023, Bisio2024, trindade2025}.
    \end{itemize}
    
    \vspace{0.4em}
    \item[3\textsuperscript{rd} Feature Engineering:] 
    Feature design varies across works:
    \begin{itemize}[leftmargin=1em, nosep, label=\textbullet]
        \item \textbf{Handcrafted features:} Statistical, spectral, and temporal descriptors informed by domain knowledge (e.g., variance, spectral centroid), offering interpretability (see Section~\ref{sec:methodology}).
        \item \textbf{Deep learning features:} CNN, RNN, and Transformer models that automatically learn hierarchical representations but reduce interpretability~\cite{Diaz2023, Li2024.1}.
        \item \textbf{Hybrid approaches:} Combine handcrafted and learned features to balance physical meaning and discriminative power~\cite{Zhang2022, Kong2022}.
    \end{itemize}

    \vspace{0.4em}
    \item[4\textsuperscript{th} Learning Model:] 
    Systems employ a range of classifiers:
    \begin{itemize}[leftmargin=1em, nosep, label=\textbullet]
        \item \textbf{Classical ML:} SVM, Random Forest, KNN, and Gaussian Naïve Bayes provide efficiency and transparency~\cite{Sobehy2020, Rocamora2020, trindade2025}.
        \item \textbf{Deep Learning:} CNNs for spatial cues, RNNs/LSTMs for temporal sequences, and Transformers for attention-based modeling~\cite{Zheng2023}.
        \item \textbf{Ensemble methods:} Multiple models combined to improve robustness and generalization~\cite{Shi2021, Lin2023, Ramesh2025}.
    \end{itemize}

    \vspace{0.4em}
    \item[5\textsuperscript{th} Evaluation Protocol:]
    Evaluation choices strongly influence reported performance:
    \begin{itemize}[leftmargin=1em, nosep, label=\textbullet]
        \item \textbf{Data splitting:} Subject-independent and session-independent splits help prevent data leakage.
        \item \textbf{Cross-validation:} K-fold protocols with within-fold normalization ensure unbiased estimates~\cite{Lin2023, trindade2025, Zhang2022}.
        \item \textbf{Metrics:} Accuracy, Precision, Recall, F1-score, ROC-AUC, per-class EER, FCS, and the GC provide complementary performance insights~\cite{Hu2024}.
    \end{itemize}
    
\end{description}

\section{Related Work}
\label{sec:related}

To facilitate systematic comparison, we present a benchmark of state-of-the-art CSI-based systems for biometric authentication, identification, and recognition. Although not all prior works explicitly target biometric authentication, many leverage Wi-Fi CSI to extract person-specific signatures that enable these tasks. 

Methodological choices fundamentally dictate adversarial resilience. While leveraging {5\,GHz and complex-domain features increases entropy to mitigate geometric spoofing, the prevailing focus on dynamic traits overlooks the physiological stability of static trust anchors. Crucially, reliance on aggregate metrics obscures ``security holes,'' masking concentrated per-user vulnerabilities that remain invisible without distributional analysis (FCS, GC).

Notably, only a small portion of the literature focuses exclusively on \textit{static CSI acquisition}~\cite{trindade2025, Ramesh2025}, as most approaches rely on dynamic movements or behavioral patterns. This scarcity constitutes a critical security blind spot: while dynamic features (e.g., gait, gestures) are inherently susceptible to behavioral mimicry and replay attacks, static acquisition isolates physiological channel modulation—governed by the user's internal dielectric properties—providing a harder-to-forge trust anchor that remains insufficiently characterized in current threat models. An analysis of Table~\ref{tab:benchmark_literature} highlights representative works across these key dimensions.

\textbf{Hardware Platform Diversity:} Early work relied predominantly on desktop-based Intel~5300 and Atheros platforms, whereas recent studies increasingly adopt embedded devices (e.g., Raspberry~Pi with Nexmon)~\cite{Turetta2022, Guan2023, trindade2025}. This shift reflects a growing trend toward practical, deployable solutions suitable for edge environments.

\textbf{Frequency Band Selection:} Most static authentication systems prefer the 5\,GHz band due to its finer spatial resolution and improved sensitivity to small-scale biometric features compared to 2.4\,GHz~\cite{Guan2023}.

\textbf{Static vs.\ Dynamic Focus:} The literature remains dominated by dynamic authentication approaches (e.g., gait, keystroke, gestures), with only a limited number of works explicitly addressing static palm-based biometrics~\cite{trindade2025, Ramesh2025}. This gap underscores the distinct relevance of our approach.

\textbf{Evaluation Metrics:} Reporting of EER and ROC-AUC remains inconsistent across studies~\cite{Kong2021, Gu2021, Gu2022, Wang2022, Kong2023, Bisio2024, Lin2023, Guan2023}, hindering direct performance comparisons. Our work addresses this deficiency through comprehensive security-oriented metrics, including per-class EER analyzed alongside FCS and GC.

\textbf{Feature Domain:} Systems that incorporate both amplitude and phase information~\cite{Kong2021, Shi2021, Zhang2022, Geng2022, Lin2023, Kong2023, Bisio2024, Li2024.2, trindade2025} consistently outperform amplitude-only methods, highlighting the critical value of full complex-domain processing.

\subsection{Identified Gaps}
\label{subsec:gaps}

We identify critical gaps limiting the scientific maturity of CSI-based biometrics, categorized into three primary dimensions.

\textbf{Reproducibility and Standardization Crisis:} The field suffers from a lack of standardized datasets and evaluation protocols (closed-world setting~\cite{Arp2022}), compounded by a heavy reliance on proprietary hardware (e.g., Intel~5300), which appears in 81.25\% of the surveyed works. This scarcity of open benchmarks hinders fair comparison and scientific reproducibility. To advance the field, wider adoption of open-source platforms (e.g., Nexmon), currently utilized in only 18.75\% of studies, along with strict documentation, is essential to overcome the limitations of black-box implementations.

\textbf{Methodological and Evaluation Deficiencies:} Methodological rigor is often lacking regarding preprocessing and metrics. Specific outlier removal techniques—such as IQR in the frequency domain and MAD in the time domain—as well as the trade-offs between interpretable handcrafted features and deep learning representations, are rarely evaluated systematically. Notably, 43.75\% of the benchmarked works neglect handcrafted features entirely. Furthermore, many studies report only aggregate accuracy, omitting critical security metrics such as EER, FAR, and FRR (absent in 62.50\% of works)—\textbf{thereby masking intrusion vulnerabilities}—and ROC-AUC (missing in 87.50\%), while failing to validate systems across diverse environments (underspecification~\cite{Jacobs2022}), which obscures true algorithmic robustness.

\textbf{Underexplored Static Modalities:} While dynamic modalities such as gait and gestures dominate the literature, static palm-based authentication remains largely underexplored, with 87.50\% of studies focusing exclusively on dynamic approaches. This gap is noteworthy because static acquisition provides stable, repeatable biometric signatures that are less sensitive to temporal behavioral fluctuations, motivating our specific focus on this domain.

\section{Methodology: A Systematic Evaluation Framework}
\label{sec:methodology}

This section systematizes evaluation practices across the CSI-based biometric literature, establishing a unified framework that exposes how methodological choices influence security properties. Rather than proposing a novel pipeline, we consolidate fragmented practices from prior work—spanning preprocessing, feature extraction, and validation protocols—to enable reproducible and security-oriented assessment. This framework serves as a lens through which we analyze the field's methodological inconsistencies and their security implications.

\begin{figure}[t!]
    \centering
    \includegraphics[width=1\linewidth, height=10cm, keepaspectratio]{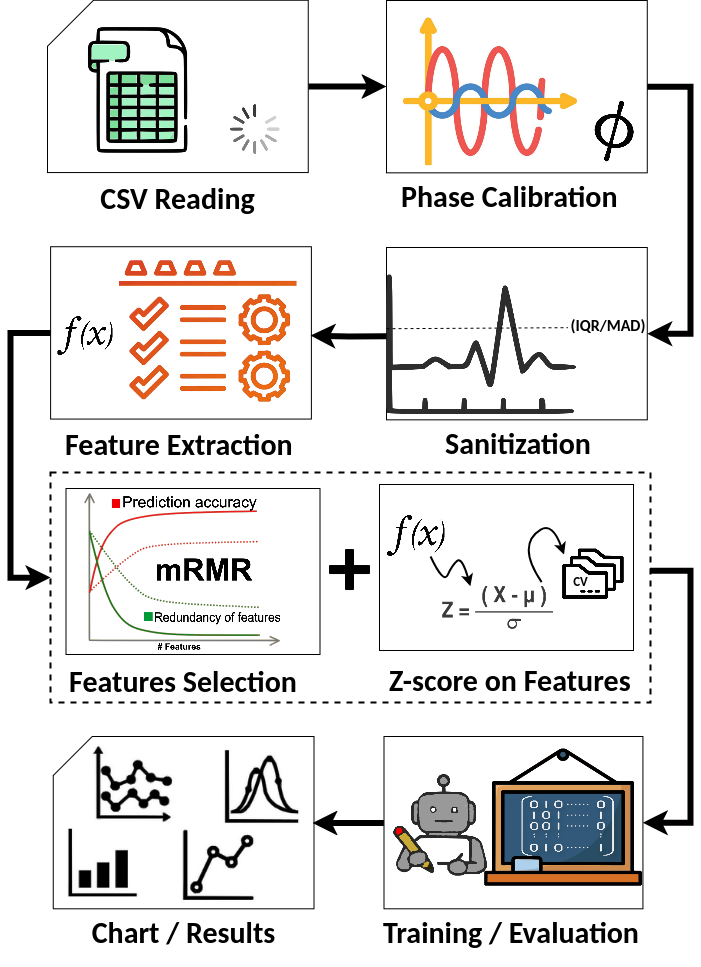}
    \caption{A Systematic Evaluation for CSI-Based Biometric Authentication.}
    \label{fig:pipeline}
\end{figure}

The systematized workflow integrates preprocessing techniques (phase calibration, outlier removal), feature extraction strategies (amplitude, phase, energy domains), and validation protocols (cross-validation, metric selection) commonly found across the literature. This design explicitly structures evaluation to guarantee reproducibility, prevent data leakage, and preserve physical interpretability—principles that are inconsistently applied in existing work. Figure~\ref{fig:pipeline} illustrates how these components interconnect in a unified evaluation architecture.

\subsection{Preprocessing and Cleaning}
\label{subsec:preprocessing}

Raw CSI readings contain measurement noise and hardware-induced distortions that vary across sensing platforms. To ensure statistical consistency and physical plausibility, \textbf{Phase Calibration} is systematically applied first, utilizing unwrapping and detrending to mitigate CFO, SFO, and PDD—artifacts that are inconsistently addressed in prior work. This ensures that the residual phase $\phi_{\text{calibrated}}$ represents true propagation effects rather than hardware drift.

Following calibration, the framework applies outlier removal techniques commonly used in the literature. \textbf{Subcarrier-level filtering} removes frequency-domain outliers via the Interquartile Range (IQR) method, discarding frequency bins with energy values outside $[Q1 - 1.5 \times IQR, Q3 + 1.5 \times IQR]$ (typically targeting border subcarriers). \textbf{Sample-level correction} identifies temporal anomalies via Median Absolute Deviation (MAD)—implemented as a rolling-window filter with a rejection threshold of $6 \times \text{MAD}$—replacing them using linear interpolation. These techniques, while individually present in some prior work, are rarely combined systematically, creating evaluation inconsistencies.

Figure~\ref{fig:Spectra} demonstrates the critical impact of systematic preprocessing. In the absence of IQR and MAD filtering, the spectrum exhibits distorted amplitudes exceeding 10 (z-score); post-filtering, values are constrained to an approximate range of $[-1,3]$. This refinement retains only legitimate channel variations induced by the hand's interaction with the EM field, rendering the CSI physically interpretable for feature extraction—a requirement that many prior evaluations fail to meet.

\begin{figure}[t!]
    \centering
    \begin{subfigure}[b]{0.48\columnwidth}
        \centering
        \includegraphics[width=\textwidth]{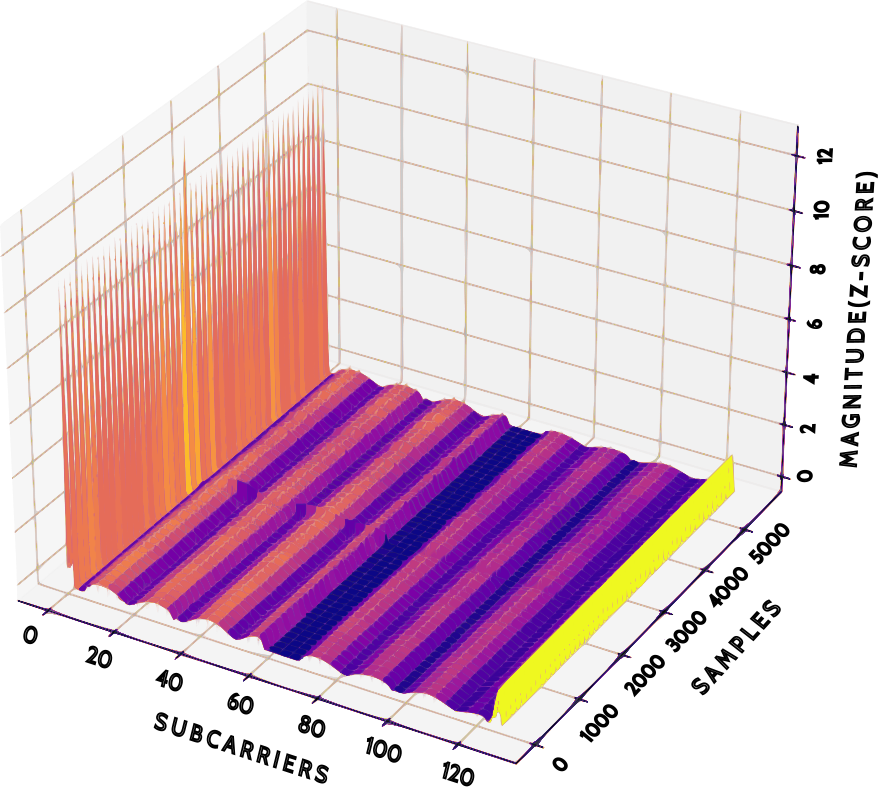}
        \caption{Unfiltered Spectrum}
        \label{fig:No-f-spec}
    \end{subfigure}
    \hfill
    \begin{subfigure}[b]{0.48\columnwidth}
        \centering
        \includegraphics[width=\textwidth]{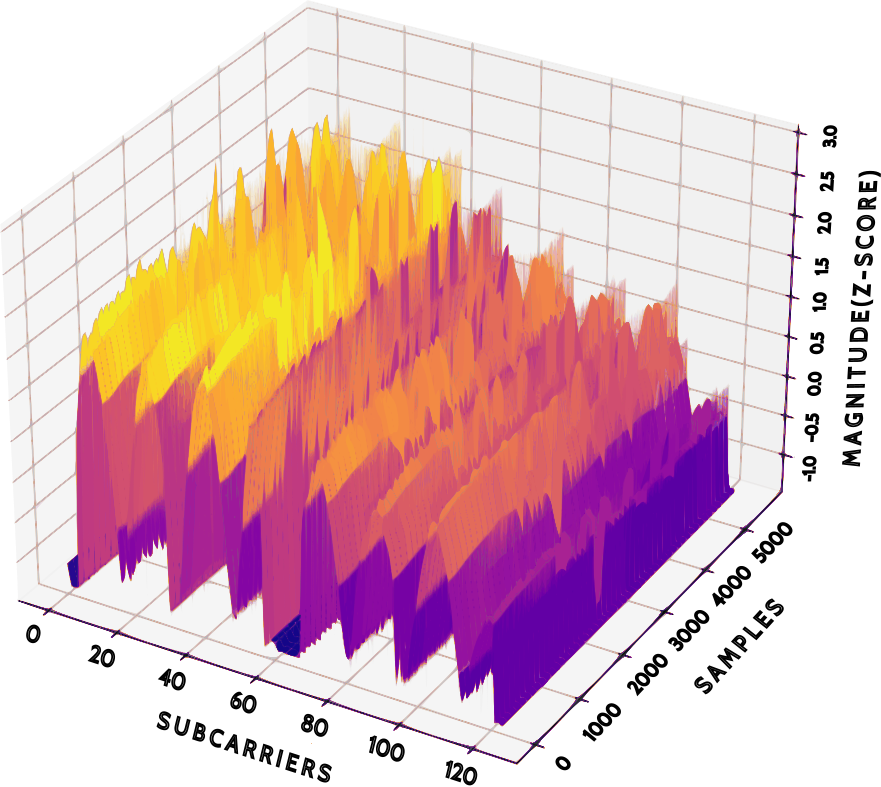}
        \caption{Filtered Spectrum}
        \label{fig:f-spec}
    \end{subfigure}
    \caption{Spectral Comparison: Unfiltered and Filtered CSI Data from a Single Subject (5-Second Acquisition).}
    \label{fig:Spectra}
\end{figure}

\subsection{Feature Extraction}
\label{subsec:feature-ext}

Feature extraction strategies across the literature vary significantly, ranging from amplitude-only approaches to complex-valued processing. Our systematization consolidates these practices into a structured taxonomy: amplitude-based, phase-based, spectral, and energy dispersion features. From each cleaned complex matrix $H(f_k, t)$, we derive descriptors that capture both large-scale (energy-level) and fine-grained (spectral-shape) variations sensitive to the micro-geometry of the human body.

\textbf{Amplitude-based features} encompass the mean, variance, skewness, and kurtosis of $|H(f_k)|$, alongside the mean subcarrier energy $E(f_k) = \frac{1}{T}\sum_{t=1}^T |H(f_k, t)|^2$. \textbf{Phase-based features} capture the mean phase $\mathbb{E}[\phi(f_k)]$, standard deviation $\sigma_{\phi}$, and phase-difference texture $\Delta \phi(f_k) = \phi(f_{k+1}) - \phi(f_k)$. \textbf{Spectral features} include the spectral centroid, flatness, and entropy derived from the 1D~FFT of the averaged magnitude response $\bar{H}(f_k)$. \textbf{Energy dispersion features} quantify power redistribution by computing ratios of reflected, absorbed, and refracted components via $\{|H|^2, \angle H\}$ statistics.

This diverse feature space reflects practices found across the literature, where different works emphasize different domains (amplitude-only vs. complex-valued). Our systematization reveals that feature choice materially influences both discriminative power and security properties—a relationship that prior work rarely analyzes explicitly.

\subsection{Evaluation Protocol and Validation}
\label{subsec:model-train}

To enable systematic comparison with prior work, we establish a unified evaluation protocol that addresses inconsistencies identified in the literature. From the five samples collected per subject (right hand), four were allocated to the training pipeline, while the remaining sample was reserved exclusively for the inference phase—a split strategy that aligns with subject-independent evaluation principles found in some prior work, but inconsistently applied across the field.

To ensure reliability and expose data leakage risks prevalent in the literature, our framework enforces strict data isolation principles. We validate leakage-free pipelines (employing feature selection and normalization within-fold) against global baselines, rejecting any configuration where the accuracy discrepancy exceeded $0.01$ (1~p.p.)—a threshold that reveals evaluation artifacts common in prior work.

To assess robustness and expose how methodological choices influence security properties, we investigate distinct evaluation configurations, specifically contrasting global versus rolling-window filtering strategies and varying input granularity (window sizes between 50 and 500 samples). These variations reveal how evaluation choices—often made arbitrarily in prior work—materially affect reported performance and security metrics.

Hyperparameter optimization follows standard practices (grid search with stratified k-fold cross-validation, $k=10$) across six classifiers commonly used in the literature: Decision Tree, SVM (RBF), KNN, Gaussian NB, Random Forest, and MLP. This systematic evaluation enables direct comparison with prior work while exposing how model choice interacts with feature selection and preprocessing.

\begin{table*}[t!]
    \centering
    \caption{Performance Comparison of Classification Models: Impact of Window Size.}
    \label{tab:model_comparison_combined}
    \renewcommand{\arraystretch}{1.2}
    \begin{minipage}[b]{0.49\textwidth}
        \centering
        \subcaption{50-Sample Window Size}
        \label{tab:results_50}
        \resizebox{\linewidth}{!}{%
            \begin{tabular}{lcccccc}
            \toprule
            \textbf{Model} & \textbf{Prec (\%)} & \textbf{Spec (\%)} & \textbf{Rec (\%)} & \textbf{F1 (\%)} & \textbf{ROC (\%)} & \textbf{EER (\%)} \\
            \midrule
            Decision Tree & 95.00 & 99.74 & 95.01 & 95.00 & 97.37 & 02.63 \\
            \textbf{SVM (RBF)}     & \textbf{99.36} & \textbf{99.97} & \textbf{99.36} & \textbf{99.36} & \textbf{99.99} & \textbf{00.19} \\
            KNN           & 97.61 & 99.87 & 97.61 & 97.60 & 99.69 & 00.60 \\
            Naive Bayes   & 64.91 & 98.08 & 63.38 & 62.14 & 95.99 & 09.01 \\
            \textbf{Random Forest} & \textbf{99.09} & \textbf{99.95} & \textbf{99.09} & \textbf{99.08} & \textbf{99.99} & \textbf{00.28} \\
            \textbf{MLP}           & \textbf{99.50} & \textbf{99.97} & \textbf{99.50} & \textbf{99.50} & \textbf{99.99} & \textbf{00.14} \\
            \bottomrule
            \end{tabular}
        }
    \end{minipage}
    \hfill
    \begin{minipage}[b]{0.49\textwidth}
        \centering
        \subcaption{500-Sample Window Size}
        \label{tab:results_500}
        \resizebox{\linewidth}{!}{%
            \begin{tabular}{lcccccc}
            \toprule
            \textbf{Model} & \textbf{Prec (\%)} & \textbf{Spec (\%)} & \textbf{Rec (\%)} & \textbf{F1 (\%)} & \textbf{ROC (\%)} & \textbf{EER (\%)} \\
            \midrule
            Decision Tree & 87.66 & 99.34 & 87.46 & 87.49 & 93.40 & 06.60 \\
            SVM (RBF)     & 95.12 & 99.74 & 95.11 & 95.04 & 99.83 & 00.95 \\
            KNN           & 92.76 & 99.61 & 92.69 & 92.57 & 98.82 & 02.11 \\
            Naive Bayes   & 69.52 & 98.28 & 67.17 & 66.08 & 96.33 & 08.18 \\
            \textbf{Random Forest} & \textbf{97.41} & \textbf{99.86} & \textbf{97.36} & \textbf{97.36} & \textbf{99.96} & \textbf{00.82} \\
            \textbf{MLP}           & \textbf{96.49} & \textbf{99.81} & \textbf{96.47} & \textbf{96.45} & \textbf{99.95} & \textbf{00.70} \\
            \bottomrule
            \end{tabular}
        }
    \end{minipage}
\end{table*}

\textbf{Standard Metrics:} Performance assessment relies on accuracy, macro-precision, recall, F1-score, and ROC-AUC—metrics that are inconsistently reported across the literature. Our systematization reveals that metric choice often obscures security-relevant properties.

\textbf{Equal Error Rate (EER):} Defined as the operating point where $FAR=FRR$, we enforce a strict maximum threshold of 5.00\% to align with security requirements. However, our analysis reveals that many prior works either omit EER entirely or report it only at the aggregate level, masking per-user vulnerabilities.

\textbf{Gini Coefficient (GC):} This metric quantifies the inequality of error distribution across the user population—a dimension that prior work almost universally ignores. A low GC value indicates uniform security, verifying that no specific subset of users is disproportionately vulnerable to false acceptances or rejections.

\textbf{Aggregated OVR FCS:} This metric visualizes global discriminability by aggregating One-vs-Rest probabilities into \textit{Genuine} ($P(U)$) and \textit{Impostor} ($P(X \neq U)$) distributions. A robust system exhibits a clear separation gap—with Genuine scores clustering near 1.0 and Impostor scores suppressed near 0.0—revealing security properties that aggregate metrics cannot capture.

This rigorous validation structure ensures that evaluation practices align with security requirements, exposing how methodological inconsistencies in prior work compromise security assessment.

\section{Results and Insights}
\label{sec:results}

This section presents insights derived from applying our unified evaluation framework to expose security-relevant properties that remain hidden under traditional reporting practices. Rather than emphasizing performance metrics, we focus on how systematic evaluation reveals structural weaknesses in current assessment methodologies. Our analysis demonstrates that common evaluation practices—especially reliance on aggregate accuracy—mask critical security vulnerabilities.

\begin{figure}[t!]
    \centering
    \includegraphics[width=\linewidth, height=4.8cm, keepaspectratio]{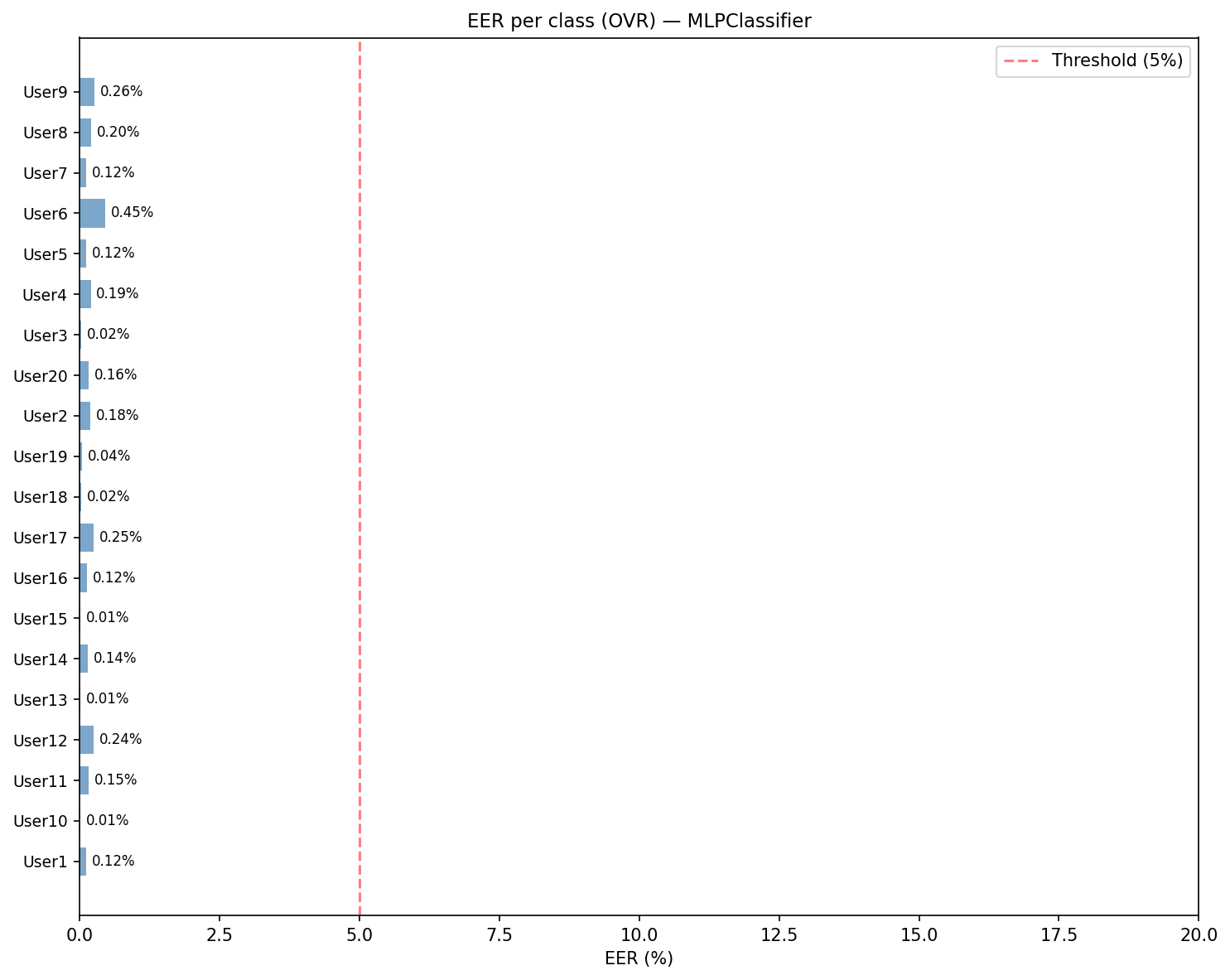}
    \vspace{0.1cm}
    \caption{Per-subject EER performance using the MLP classifier.}
    \label{fig:EER}
\end{figure}

\subsection{Quantitative Evaluation}
\label{subsec:evaluate}

Our systematic evaluation reveals that inter-user CSI variability encodes unique amplitude and phase signatures, as evidenced by distinct separation in confusion matrices. However, this analysis also exposes a critical limitation: aggregate metrics mask significant per-user variability that has security implications.

The evaluation framework demonstrates how methodological choices—specifically window size and feature selection—materially influence both aggregate performance and security properties. As detailed in Table~\ref{tab:model_comparison_combined}, smaller window sizes (50 samples) yield higher aggregate metrics, but this observation alone is insufficient for security assessment.

The framework reveals that smaller window sizes provide more feature instances, but this observation must be contextualized: the security implications of this choice extend beyond aggregate accuracy. Our analysis demonstrates that evaluation granularity choices—often made without security consideration—affect both performance and vulnerability profiles.

\begin{figure}[t!]
    \centering
    \includegraphics[width=\linewidth, height=4.87cm, keepaspectratio]{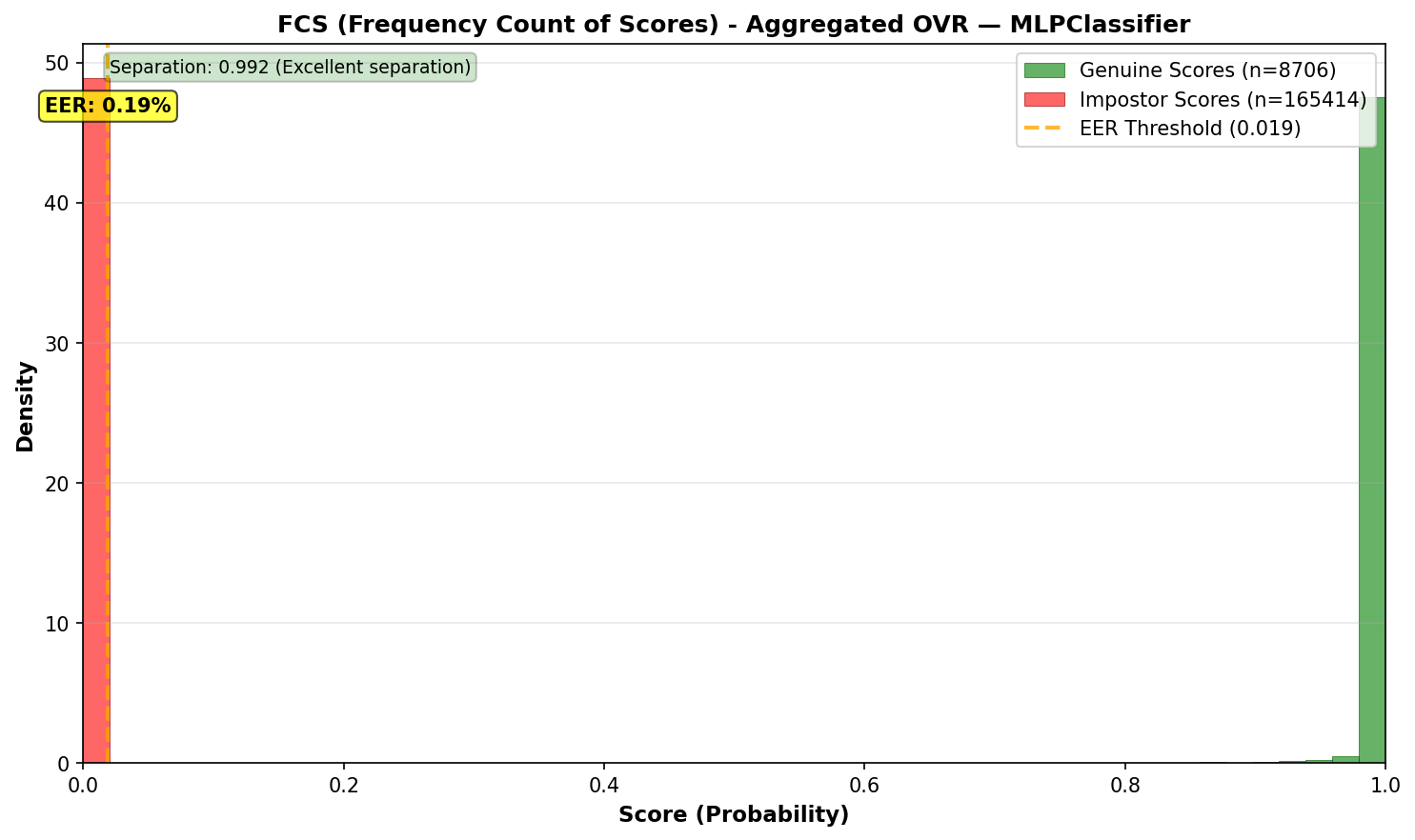}
    \caption{Probability score distribution produced by the MLP model.}
    \label{fig:FCS}
\end{figure}

\subsection{Security Metrics Results}
\label{subsec:sec-metric-disc}

This analysis shifts to security indicators that prior work systematically omits. Our unified framework exposes how aggregate metrics mask critical vulnerabilities. For instance, while some models achieve low aggregate EER, per-user analysis reveals extreme threshold variability that compromises security for specific individuals—a finding that aggregate reporting cannot capture (see Figure~\ref{fig:EER}).

The framework demonstrates that models optimized for aggregate precision exhibit significant inter-class variability. Some models achieve high separation indices (see Figure~\ref{fig:FCS} approaching 1.0) but operate with extreme decision thresholds (ranging from $\approx 0.008$ to $\approx 0.987$), resulting in local overlaps between genuine and impostor score distributions. For affected users, this creates a critical trade-off, forcing the system to compromise either security (high FAR) or usability (high FRR)—a vulnerability that aggregate metrics completely obscure.

\begin{setupbox}
    \textbf{Key Insight:} Our systematization reveals a fundamental trade-off between aggregate precision and distributional robustness. Models that optimize for aggregate metrics (e.g., minimal EER) may concentrate security risks on a minority of users, while models that prioritize distributional uniformity sacrifice marginal aggregate gains for equitable protection. This trade-off is invisible under traditional reporting practices, highlighting the critical need for distributional security metrics.
\end{setupbox}

As detailed in Table~\ref{tab:direct_comparison}, despite some models exhibiting lower mean EER, they compromise usability and security for a significant portion of the user base due to volatile operational thresholds. Specifically, a substantial fraction of subjects operate at extreme thresholds ($<0.1$ or $>0.7$), indicating potential security vulnerabilities or usability issues, while only a minority achieve ideal threshold ranges (0.1--0.7).

Other models demonstrate intermediate stability, with a smaller fraction of subjects at extreme thresholds. While this represents improvement, it still falls short of the stability required for production deployment.

In contrast, models that prioritize distributional robustness ensure that the majority of subjects operate at optimal, balanced thresholds, with only a small fraction at extreme values. This provides uniform security and usability across the vast majority of the population, satisfying stability requirements for production deployment.

\begin{table}[t!]
    \centering
    \caption{Stability Comparison: MLP and Random Forest.}
    \label{tab:direct_comparison}
    \resizebox{\columnwidth}{!}{%
    \begin{tabular}{lccc}
        \toprule
        \textbf{Metric / Aspect} & \textbf{MLP} & \textbf{Random Forest} & \textbf{Winner} \\
        \midrule
        \textbf{Threshold Variability} & 0.3503 & 0.0877 & \textcolor{green}{\boldmath$\checkmark$} RF (4$\times$ lower) \\
        \textbf{Extreme Thresholds} & 14/20 (70\%) & 2/20 (10\%) & \textcolor{green}{\boldmath$\checkmark$} RF \\
        \textbf{Ideal Thresholds} & 6/20 (30\%) & 18/20 (90\%) & \textcolor{green}{\boldmath$\checkmark$} RF \\
        \textbf{Consistency} & \textcolor{red}{\textbf{$\times$}} Low & \textcolor{green}{\boldmath$\checkmark$} High & \textcolor{green}{\boldmath$\checkmark$} RF \\
        \textbf{Uniform Security} & \textcolor{red}{\textbf{$\times$}} No & \textcolor{green}{\boldmath$\checkmark$} Yes & \textcolor{green}{\boldmath$\checkmark$} RF \\
        \textbf{Production Readiness} & \textcolor{red}{\textbf{$\times$}} No & \textcolor{green}{\boldmath$\checkmark$} Yes & \textcolor{green}{\boldmath$\checkmark$} RF \\
        \bottomrule
    \end{tabular}
    }
\end{table}

\subsection{Feature Importance Analysis}
\label{subsec:feature-importance}

To identify the most discriminative descriptors for CSI-based authentication, we applied feature selection criteria (mRMR) to the set of extracted features, reducing dimensionality systematically. This analysis reveals patterns that inform feature engineering practices across the literature.

The analysis reveals that spectral and energy-based descriptors dominate the feature landscape, accounting for the majority of top-ranked selections. Specifically, spectral centroid, energy reflection metrics, and energy skewness achieved the highest scores, indicating strong discriminative power. These features effectively capture how the user's hand modifies multipath propagation—via absorption, reflection, and scattering—across different subcarrier frequencies (see Figure~\ref{fig:feature-import}).

Notably, raw phase-based features were absent from the top rankings. This suggests that global phase information is insufficient without proper calibration—a finding that challenges amplitude-only approaches while highlighting the importance of systematic preprocessing.

A critical insight from this analysis lies in the role of energetic features. Based on the Poynting theorem, we define the spectral energy as $E(f_k)=|H(f_k)|^2$. From this, two distinct categories of descriptors emerge: statistical energy features (skewness, kurtosis, entropy) that quantify distribution, and empirical energy features that decompose the signal into physical interactions (reflected, absorbed, refracted). The high ranking of empirical features highlights the biometric validity of physical interactions—an insight that prior work rarely articulates explicitly.

\begin{figure}[t!]
    \centering
    \includegraphics[width=\linewidth, keepaspectratio]{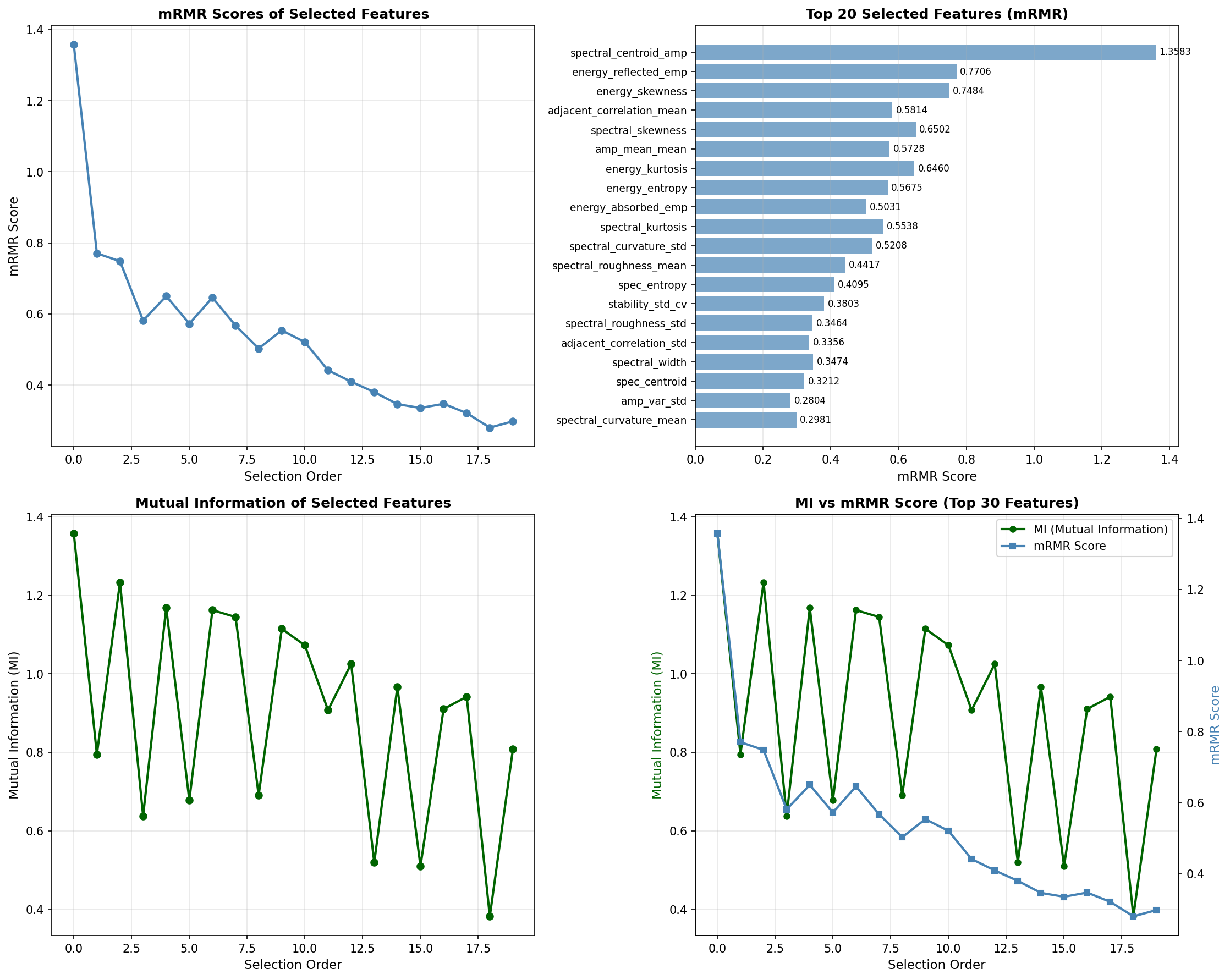}
    \caption{Ranking of the top 20 discriminative features selected using Mutual Information and the mRMR score.}
    \label{fig:feature-import}
\end{figure}

These two groups exhibit a complementary relationship: statistical features describe the \textit{distribution} of energy, whereas empirical features describe the \textit{fate} of that energy (interaction). Coupled with spectral descriptors, the selected subset provides a physically interpretable representation of hand-induced CSI variations, merging statistical uniqueness with biomechanical meaning—a balance that prior work achieves inconsistently.

\subsection{Gini Coefficient: A Complementary Metric for Evaluation}
\label{subsec:complement-metric}

\subsubsection*{Stability Analysis: Aggregate vs. Distributional Metrics}
While some models exhibit superior aggregate metrics—achieving minimal EER and near-perfect ROC-AUC—analysis via distributional metrics exposes significant inter-class variability. Models optimized for aggregate precision operate with extreme decision thresholds, resulting in local overlaps between genuine and impostor score distributions. For affected users, this creates a critical trade-off, forcing the system to compromise either security (high FAR) or usability (high FRR).

To quantify this disparity, we employ the Gini Coefficient, which measures the inequality of the error distribution across the user population at the EER threshold~\cite{hong2021gini}. As validated in Table~\ref{tab:FCS-GC_mlp-rf}, a high GC value indicates that errors are not randomly distributed but rather concentrated on a specific subset of users—a vulnerability that aggregate metrics completely mask.

Notably, despite excellent average performance, models optimized for aggregate precision demonstrate significant instability. High overall inequality (GC exceeding 45\%) reveals ``security holes'' for specific individuals. Similarly, some models exhibit the most severe concentration of security risks in specific error types (FAR or FRR), reaching GC values exceeding 60\%. This confirms that optimization for aggregate precision may exacerbate per-user disparity—a finding that challenges common evaluation practices.

In stark contrast, models that prioritize distributional robustness maintain substantially lower inequality indices (GC below 40\%). By distributing errors more evenly, they ensure a democratic and predictable security level across the entire population, avoiding the extreme vulnerabilities observed in aggregate-optimized architectures.

\begin{table}[t!]
    \centering
    \caption{Comparison of FAR, FRR, and Gini Coefficient Metrics Across Classifiers.}
    \label{tab:FCS-GC_mlp-rf}
    \resizebox{\columnwidth}{!}{%
    \begin{tabular}{lccccc}
        \toprule
        \textbf{Model} & \textbf{FAR (\%)} & \textbf{FRR (\%)} & \textbf{GC$_{\text{FAR}}$ (\%)} & \textbf{GC$_{\text{FRR}}$ (\%)} & \textbf{GC$_{\text{mean}}$ (\%)} \\
        \midrule
        DecisionTree  & 0.26 & 4.99 & 35.96 & 36.95 & 36.46 \\
        \textbf{SVM (RBF)}     & \textbf{0.03} & \textbf{0.64} & \textbf{61.15} & \textbf{35.49} & \textbf{48.32} \\
        KNN           & 0.13 & 2.39 & 46.41 & 46.28 & 46.34 \\
        GaussianNB    & 1.92 & 36.62 & 39.38 & 39.56 & 39.47 \\
        \textbf{RandomForest} & \textbf{0.05} & \textbf{0.91} & \textbf{39.68} & \textbf{39.53} & \textbf{39.61} \\
        \textbf{MLPClassifier} & \textbf{0.03} & \textbf{0.50} & \textbf{55.27} & \textbf{41.43} & \textbf{48.35} \\
        \bottomrule
    \end{tabular}%
    }
\end{table}

\section{Discussion}
\label{sec:discussion}

This section synthesizes insights from our systematization, assessing how methodological inconsistencies in the literature compromise security evaluation and identifying critical lessons for the community.

\subsection{Strengths and Limitations}
\label{subsec:limitations}

Our systematization reveals four critical dimensions that influence security properties in CSI-based biometrics:

\textbf{Reproducibility and Accessibility:} The field's heavy reliance on proprietary hardware (e.g., Intel 5300) in the majority of prior work creates reproducibility barriers. Our analysis demonstrates that commodity hardware (e.g., Raspberry Pi/Nexmon) enables reproducible evaluation, but this shift requires systematic adoption across the community to enable fair comparison.

\textbf{Feature Engineering Insights:} Our systematization confirms that spectral and empirical energy features dominate discriminative power, while raw phase features require systematic calibration. This finding challenges amplitude-only approaches prevalent in some prior work while highlighting the importance of systematic preprocessing—a practice inconsistently applied across the literature.

\textbf{Distributional Robustness:} Our analysis reveals a critical trade-off between aggregate precision and stability that prior work systematically ignores. Models optimized for aggregate metrics may concentrate security risks on a minority of users, while models prioritizing distributional uniformity ensure equitable protection—a dimension that aggregate reporting cannot capture.

\textbf{Dual-Domain Sensing:} Systematic phase calibration enables exploitation of the full complex-valued channel response, capturing richer biometric signatures than amplitude-only approaches. However, our systematization reveals that phase processing is inconsistently applied, with many works either omitting phase entirely or applying calibration inconsistently.

However, limitations persist. The \textbf{closed-world evaluation} subjects models to the \textbf{underspecification problem}~\cite{Jacobs2022}, potentially limiting robustness against unmodeled environmental variations—a limitation that affects the entire field, not just our evaluation. Furthermore, the current \textbf{absence of drift adaptation} in the literature restricts resilience against long-term temporal fluctuations or novel anomalies.

\subsection{Security and Privacy Implications}
\label{subsec:s&c-implications}

Our systematization exposes how evaluation practices in the literature compromise security assessment:

\textbf{Spoofing Resistance:} Unlike surface-level biometrics (e.g., fingerprints), CSI signatures are modulated by internal tissue properties (dielectric constants of bone, blood, and muscle). This physical dependency provides inherent resistance against geometric spoofing and simple replay attacks. However, our analysis reveals that many prior works fail to evaluate spoofing resistance explicitly, creating security assessment gaps.

\textbf{Privacy-Preserving Architecture:} To mitigate surreptitious tracking risks, edge-processing architectures where feature extraction occurs locally are essential. Transmitting only anonymized, non-invertible feature vectors prevents reconstruction of raw physiological data. However, our systematization reveals that privacy implications are rarely addressed in prior work, creating deployment risks.

\textbf{Metric-Driven Auditing:} Relying solely on aggregate metrics (EER) is insufficient for security certification. As demonstrated by our analysis, excellent average performance can mask ``security holes'' for minority users. Distributional metrics like the Gini Coefficient must be standard requirements to certify equitable protection—a practice that prior work almost universally omits.

\begin{table}[t!]
    \centering
    \caption{Statistical Reliability Analysis: EER, Uncertainty, and CI Width (BioQuake Components).}
    \label{tab:bioquake_metrics}
    \resizebox{\columnwidth}{!}{%
    \begin{tabular}{lccc}
        \toprule
        \textbf{Model} & \textbf{EER (\%)} & \textbf{Uncertainty (\%)} & \textbf{CI Width (\%)} \\
        \midrule
        DecisionTree    & 02.63 & 01.75 & 01.64 \\
        \textbf{SVM (RBF)}       & \textbf{00.19} & \textbf{00.15} & \textbf{00.14} \\
        KNN             & 00.60 & 00.51 & 00.48 \\
        GaussianNB      & 09.01 & 06.45 & 06.03 \\
        \textbf{RandomForest}    & \textbf{00.28} & \textbf{00.21} & \textbf{00.20} \\
        \textbf{MLPClassifier} & \textbf{00.14} & \textbf{00.11} & \textbf{00.10} \\
        \bottomrule
    \end{tabular}%
    }
\end{table}

\section{Future Directions}
\label{sec:future}

This study establishes a baseline for static palm-based CSI authentication, yet several avenues remain for enhancing system resilience and evaluation standards.

We propose that the \textbf{BioQuake} metric (see Figure~\ref{fig:bioquake})~\cite{Al-Refai2025, Fallahi2025} should effectively supersede the traditional EER in future evaluations. As detailed in Table~\ref{tab:bioquake_metrics}, BioQuake inherently computes the EER while augmenting it with critical statistical context: the standard deviation (Uncertainty) and the Confidence Interval (CI) Width. This provides a multidimensional view of model reliability that a single scalar value cannot convey. However, BioQuake must not be used in isolation. To ensure robust security, it is imperative to integrate it with state-of-the-art distributional metrics, specifically the \textbf{Feature Consistency Score (FCS)} and the \textbf{Gini Coefficient (GC)}. Optimal evaluation frameworks must adopt a hybrid approach: utilizing BioQuake to assess statistical \textit{confidence} and FCS/GC to quantify the \textit{uniformity} of protection, ensuring democratic security for all users.

\subsection{Evolutionary Model Optimization}
\label{subsec:optimization}
While Random Forest demonstrated superior distributional stability in our tests, its performance relies heavily on hyperparameter tuning. Recent advancements suggest that evolutionary algorithms, such as \textbf{Bitterling Fish Optimization (BFO)}, can significantly enhance CSI-based sensing~\cite{Wang2024.1}. By automating the selection of optimal parameters (e.g., number of trees, split criteria), BFO-RF frameworks have shown improved robustness against environmental noise and higher generalization capabilities compared to standard grid-search methods. Future work should investigate BFO-driven optimization to further minimize the trade-off between aggregate precision and per-user stability in biometric authentication.

\subsection{Anomaly Detection and Drift Compensation}
\label{subsec:anomaly-drift}

Future research must also address the dynamic nature of wireless environments through two critical mechanisms:

\textbf{Context-Aware Anomaly Detection:} To differentiate legitimate biometric variations from external interference (e.g., electromagnetic noise or far-field scattering from unauthorized individuals~\cite{Li2024}), the system requires unsupervised learning modules. Techniques like Autoencoders or One-Class SVMs can effectively filter these environmental perturbations, ensuring authentication decisions remain grounded in physiological traits rather than transient noise.

\textbf{Drift as a Security Trigger:} Instead of treating failures caused by data drift (environmental or hardware aging~\cite{Kayano2025}) as simple rejections, they should function as active security signals. By correlating high BioQuake uncertainty with potential drift, the system can trigger fallback protocols—such as step-up authentication—transforming passive degradation into a proactive fraud prevention mechanism that maintains usability while alerting users to potential risks.

\begin{figure}[t!]
    \centering
    \includegraphics[width=\linewidth, keepaspectratio]{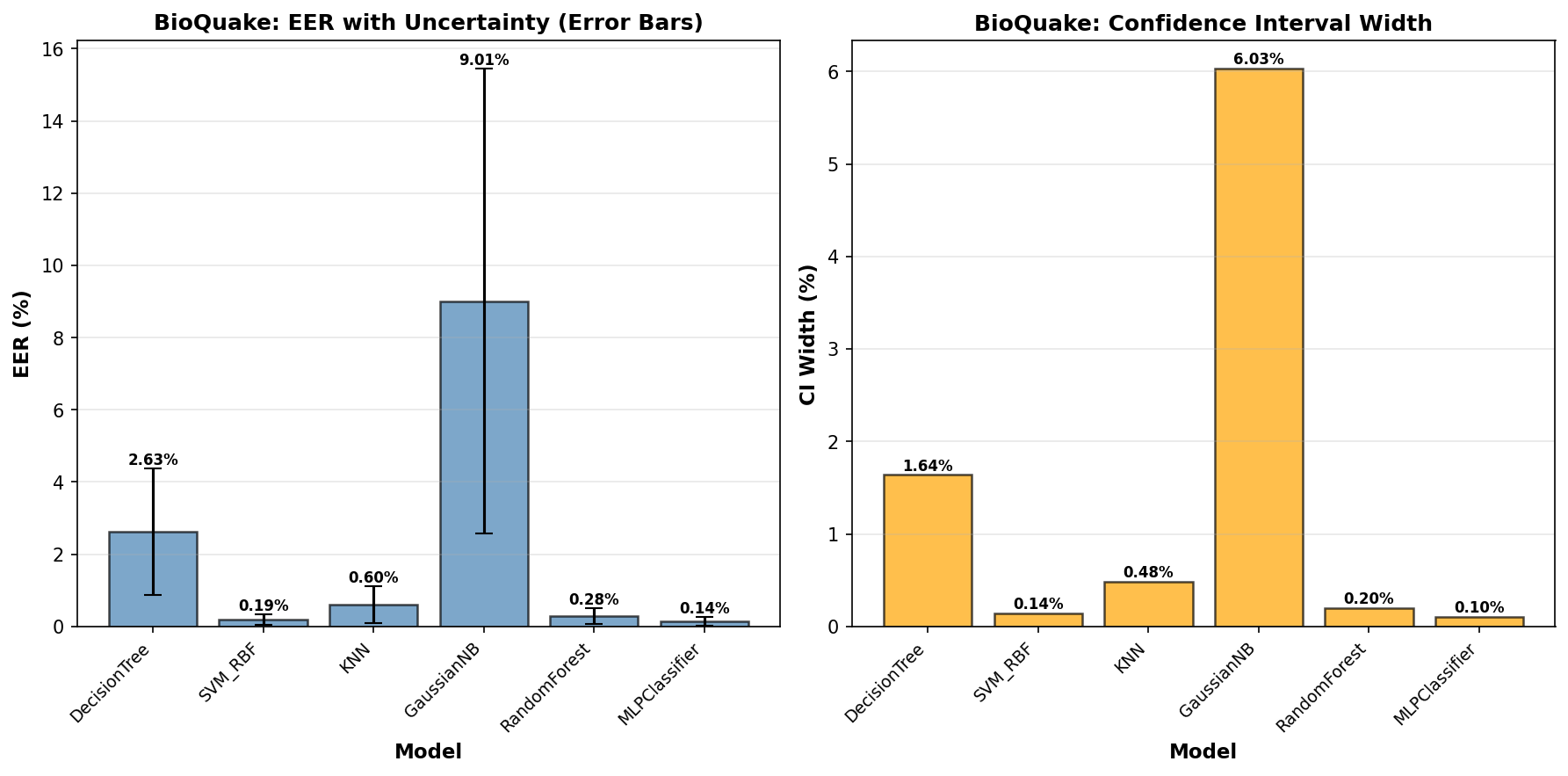}
    \caption{Breakdown of BioQuake metric components (EER, Uncertainty, CI Width) for the evaluated machine learning models.}
    \label{fig:bioquake}
\end{figure}

\subsection{IEEE 802.11bf Standardization}
\label{subsec:80211bf-standard}

The IEEE 802.11bf amendment institutionalizes Wireless WLAN Sensing (SENS), transitioning CSI acquisition from proprietary hacks to open MAC-layer signaling. By establishing vendor-agnostic protocols, it renders CSI data interoperable and machine-readable across heterogeneous devices.

\subsubsection{Compatibility and Implementation}
\label{subsubsec:80211bf-implement}

Backward compatibility with 802.11ax/be is maintained via coexistence frames. The standard defines \textit{passive} (ambient traffic) and \textit{active} (dedicated null packet) modes, while formalizing calibration procedures and stability metrics~\cite{IEEE80211bf2025} to mitigate the phase drift issues prevalent in commodity hardware.

\subsubsection{Research and Application Impact}
\label{subsubsec:80211bf-impact}

The institutionalization of CSI access accelerates innovation by ensuring consistent subcarrier reporting and phase fidelity.

\begin{setupbox}
\textbf{Impact on Sensing Domains:} Standardized CSI facilitates \textbf{RF-based Biometrics} by improving cross-platform repeatability for palm/gait recognition; enhances \textbf{Vital-Sign Monitoring} through reliable micro-Doppler detection; and enables low-latency \textbf{Real-Time Security} for intrusion detection. Furthermore, the unified sensing API bridges the gap between experimentation and deployment, allowing lightweight AI/ML models to be embedded directly at the firmware level.
\end{setupbox}

\subsubsection{Future Directions with 802.11bf}
\label{subsubsec:future-80211bf}

While establishing a foundational framework, the standard delineates critical open challenges for the research community.

\begin{setupbox}
\textbf{Key Challenges:} Future work must address \textbf{Privacy and Anonymization} by developing edge-based obfuscation techniques that sanitize data without destroying biometric utility. Additionally, high-throughput streams require efficient \textbf{Compression Codecs}, while \textbf{Spectral Expansion} to 60 GHz is needed for mm-level resolution. Finally, \textbf{Distributed Intelligence} protocols must be integrated to enable federated learning across multiple access points.
\end{setupbox}

\section{Conclusion}
\label{sec:conclusion}

This Systematization of Knowledge examined Wi-Fi CSI biometrics from a security perspective and identified structural inconsistencies that shape the field's evaluation practices. Our taxonomy organizes sensing infrastructures, signal representations, feature pipelines, learning models, and evaluation protocols, clarifying how methodological variability affects robustness and attack exposure. Using a unified framework, we showed that common reporting practices—especially reliance on aggregate accuracy—mask security-relevant effects such as uneven per-user risk and inconsistent behavior under realistic perturbations. Metrics such as per-class EER, Frequency Count of Scores (FCS), and the Gini Coefficient reveal concentrated vulnerabilities that remain hidden under conventional analyses.

These findings indicate that CSI-based biometrics cannot be meaningfully assessed without explicit adversarial models, reproducible pipelines, and evaluation protocols that reflect operational threats including replay attempts, geometric mimicry, and controlled environmental manipulation. The SoK highlights open challenges: cross-environment generalization, drift over time, limited dataset diversity, and the absence of standardized benchmarks. By consolidating methodological patterns and exposing their security implications, this work provides a foundation for more rigorous and comparable studies of CSI biometrics and delineates the boundaries within which they can be responsibly considered as authentication primitives.

\section*{Acknowledgments}
\label{sec:ack}

Removed for double-blind review.

\bibliographystyle{IEEEtran}
\bibliography{sok_wifi_csi}

\appendices
\section{Feature Extraction and Analysis}
\label{app:feature-extraction}


\textbf{Features Derived from the Average Amplitude:} 
The standard deviation and variance metrics for the amplitude are computed as follows:

\begin{equation}
\begin{split}
    \text{amp\_mean\_std} = & \\
    \sqrt{\frac{1}{K-1} \sum_{k=1}^{K} \left(\frac{1}{T} \sum_{t=1}^{T} |H(f_k, t)| - \text{amp\_mean}\right)^2}
\end{split}
\label{eq:amp-mean-std}
\end{equation}

\begin{equation}
\begin{split}
    \text{amp\_var\_mean} = & \\
    \frac{1}{K} \sum_{k=1}^{K} \frac{1}{T-1} \sum_{t=1}^{T} \left(|H(f_k, t)| - \frac{1}{T} \sum_{t'=1}^{T} |H(f_k, t')|\right)^2
\end{split}
\label{eq:amp-var-mean}
\end{equation}

\begin{equation}
\begin{split}
    \text{amp\_var\_std} = & \\
    \sqrt{\frac{1}{K-1} \sum_{k=1}^{K} \left(\text{Var}_k\{|H(f_k, t)|\} - \text{amp\_var\_mean}\right)^2}
\end{split}
\label{eq:amp-var-std}
\end{equation}

\textbf{Amplitude Statistical Moments:} The skewness of the amplitude distribution across subcarriers is defined as:
\begin{equation}
\begin{split}
    \text{amp\_skew\_mean} = 
    \frac{1}{K} \sum_{k=1}^{K} \frac{\mathbb{E}\left[\left(|H(f_k, t)| - \mu_k\right)^3\right]}{\sigma_k^3}
\end{split}
\label{eq:skew_amp}
\end{equation}

Likewise, the fourth moment of the amplitude distribution (kurtosis) across subcarriers is computed as:
\begin{equation}
\begin{split}
    \text{amp\_kurt\_mean} = & \\
    \frac{1}{K} \sum_{k=1}^{K} \frac{\mathbb{E}\left[\left(|H(f_k, t)| - \mu_k\right)^4\right]}{\sigma_k^4} - 3
\end{split}
\label{eq:kur_amp}
\end{equation}

\textbf{Phase Texture Features:} The phase gradient, defined as the difference between adjacent phases, is given by:
\begin{equation}
    \Delta\phi(f_k, t) = \phi(f_{k+1}, t) - \phi(f_k, t)
    \label{eq:phase-grad}
\end{equation}
The mean and standard deviation of the phase-difference standard deviations are computed as follows:

\begin{equation}
\begin{split}
    \text{dphi\_std\_mean} = & \\
    \frac{1}{K-1} \sum_{k=1}^{K-1} \sqrt{\frac{1}{T-1} \sum_{t=1}^{T}
    \left(\Delta\phi(f_k, t) - \frac{1}{T} \sum_{t'=1}^{T} \Delta\phi(f_k, t')\right)^2}
\end{split}
\label{eq:dphi_std_mean}
\end{equation}

\begin{equation}
\begin{split}
    \text{dphi\_std\_std} = & \\
    \sqrt{\frac{1}{K-2} \sum_{k=1}^{K-1} \left(\text{Std}_t\{\Delta\phi(f_k, t)\} - \text{dphi\_std\_mean}\right)^2}
\end{split}
\label{eq:dphi_std_std}
\end{equation}

\textbf{Energy Distribution Features:} Analysis of the energy distribution across subcarriers then enables high-granularity user differentiation~\cite{Shi2017, Mohammed2019}.

\begin{equation}
    \text{energy\_skewness} = \frac{\mathbb{E}\left[\left(E(f_k) - \mu_E\right)^3\right]}{\sigma_E^3}
    \label{eq:energy-skewness}
\end{equation}

\begin{equation}
    \text{energy\_kurtosis} = \frac{\mathbb{E}\left[\left(E(f_k) - \mu_E\right)^4\right]}{\sigma_E^4} - 3
    \label{eq:energy-kurtosis}
\end{equation}

\begin{equation}
    \text{energy\_entropy} = -\sum_{k=1}^{K} p(f_k) \log_2 p(f_k)
    \label{eq:energy-entropy}
\end{equation}
where $p(f_k) = \frac{E(f_k)}{\sum_{k'=1}^{K} E(f_{k'})}$ is the normalized energy distribution.

\textbf{Spectral Feature Decomposition:} The spectral centroid represents the center of mass of the magnitude spectrum.
\begin{equation}
    \text{spec\_centroid} = 
    \frac{\sum_{k=1}^{K} f_k \cdot |H(f_k)|}{\sum_{k=1}^{K} |H(f_k)|}
    \label{eq:spec-centroid}
\end{equation}

The spectral entropy quantifies the uniformity of the magnitude spectrum.
\begin{equation}
\begin{split}
    \text{spec\_entropy} = & \\
    -\sum_{k=1}^{K} 
    \frac{|H(f_k)|}{\sum_{k'=1}^{K} |H(f_{k'})|}
    \log_2\!\left(
    \frac{|H(f_k)|}{\sum_{k'=1}^{K} |H(f_{k'})|}
    \right)
\end{split}
\label{eq:spec-entropy}
\end{equation}

Spectral flatness measures the degree to which the spectrum resembles white noise.
\begin{equation}
\begin{split}
    \text{spec\_flatness} = 
    \frac{\sqrt[K]{\prod_{k=1}^{K} |H(f_k)|}}
    {\frac{1}{K}\sum_{k=1}^{K} |H(f_k)|}
\end{split}
\label{eq:spec-flatness}
\end{equation}

The spectral centroid for amplitude analysis is defined as:
\begin{equation}
\begin{split}
    \text{spectral\_centroid\_amp} = 
    \frac{\sum_{k=1}^{K} k \cdot |H(f_k)|}{\sum_{k=1}^{K} |H(f_k)|}
\end{split}
\label{eq:spec-centroid-amp}
\end{equation}

The spectral width, corresponding to the second moment, is given by:
\begin{equation}
\begin{split}
    \text{spectral\_width} = & \\
    \sqrt{
    \frac{
    \sum_{k=1}^{K} (k - \text{spectral\_centroid\_amp})^2 \cdot |H(f_k)|
    }{
    \sum_{k=1}^{K} |H(f_k)|
    }}
\end{split}
\label{eq:spec-width}
\end{equation}

\textbf{Empirical Energy:} Energy above the mean represents reflected components.
\begin{equation}
    R_{\text{emp}} = \frac{\mathbb{E}[E(f_k) \mid E(f_k) \geq \mu_E]}{\mu_E}
    \label{eq:reflect-energy}
\end{equation}
where $\mu_E = \frac{1}{K} \sum_{k=1}^{K} E(f_k)$ is the mean energy. Energy below the mean represents absorbed components.
\begin{equation}
    A_{\text{emp}} = \frac{\mathbb{E}[E(f_k) \mid E(f_k) < \mu_E]}{\mu_E}
    \label{eq:absorv-energy}
\end{equation}

Phase variation represents refracted or dispersed energy.
\begin{equation}
    T_{\text{emp}} = \frac{\mathbb{E}[\sigma_{\phi}(f_k)]}{\pi}
    \label{eq:refract-energy}
\end{equation}
where $\sigma_{\phi}(f_k)$ is the standard deviation of phase across time for subcarrier $k$. The normalized empirical energy features are defined as:

\begin{equation}
\begin{split}
    \text{energy\_reflected\_emp} = 
    \frac{R_{\text{emp}}}{R_{\text{emp}} + A_{\text{emp}} + T_{\text{emp}}}
\end{split}
\label{eq:emp-norm-refl}
\end{equation}

\begin{equation}
\begin{split}
    \text{energy\_absorbed\_emp} = 
    \frac{A_{\text{emp}}}{R_{\text{emp}} + A_{\text{emp}} + T_{\text{emp}}}
\end{split}
\end{equation}

\begin{equation}
\begin{split}
    \text{energy\_refracted\_emp} = 
    \frac{T_{\text{emp}}}{R_{\text{emp}} + A_{\text{emp}} + T_{\text{emp}}}
\end{split}
\label{eq:emp-norm-refr}
\end{equation}

\textbf{Temporal Variability Analysis:} Temporal variability quantifies fluctuations in CSI magnitude over time.

\begin{equation}
\begin{split}
    \text{temporal\_variability\_mean} = & \\
    \frac{1}{K} \sum_{k=1}^{K}
    \sqrt{\frac{1}{T-1} \sum_{t=1}^{T}
    \left(|H(f_k, t)| - \frac{1}{T} \sum_{t'=1}^{T} |H(f_k, t')|\right)^2}
\end{split}
\label{eq:temp-variat-mean}
\end{equation}

The standard deviation of temporal variability is:
\begin{equation}
\begin{split}
    \text{temporal\_variability\_std} = & \\
    \sqrt{\frac{1}{K-1} \sum_{k=1}^{K}
    \left(\text{Std}_t\{|H(f_k, t)|\} - \text{temporal\_variability\_mean}\right)^2}
\end{split}
\label{eq:temp-variat-std}
\end{equation}

The coefficient of variation is given by:
\begin{equation}
\begin{split}
    \text{temporal\_variability\_cv} = & \\
    \frac{\text{temporal\_variability\_mean}}
    {\frac{1}{K} \sum_{k=1}^{K} \frac{1}{T} \sum_{t=1}^{T} |H(f_k, t)|}
\end{split}
\label{eq:temp-variat-cv}
\end{equation}

\textbf{Stability Analysis:} The coefficient of variation for the stability measurement can be calculated using the equation below:
\begin{equation}
    \text{CV}_k = \frac{\text{Std}_t\{|H(f_k, t)|\}}{\mathbb{E}_t[|H(f_k, t)|]}
    \label{eq:stab-measure}
\end{equation}

The mean coefficient of variation:
\begin{equation}
    \text{stability\_mean\_cv} = \frac{1}{K} \sum_{k=1}^{K} \text{CV}_k
    \label{eq:coeff-variat}
\end{equation}

The standard deviation of coefficient of variation:
\begin{equation}
\begin{split}
    \text{stability\_std\_cv} = 
    \sqrt{\frac{1}{K-1} \sum_{k=1}^{K} \left(\text{CV}_k - \text{stability\_mean\_cv}\right)^2}
\end{split}
\label{eq:coeff-variat-std}
\end{equation}

\textbf{Adjacent Subcarrier Correlation:} The correlation between adjacent subcarriers is defined as:
\begin{equation}
    \rho_k = 
    \frac{\text{Cov}(|H(f_k, t)|, |H(f_{k+1}, t)|)}
    {\sqrt{\text{Var}(|H(f_k, t)|)\,\text{Var}(|H(f_{k+1}, t)|)}}
    \label{eq:adj-subc-corr}
\end{equation}

The mean adjacent correlation is given by:
\begin{equation}
\begin{split}
    \text{adjacent\_correlation\_mean} = 
    \frac{1}{K-1} \sum_{k=1}^{K-1} \rho_k
\end{split}
\end{equation}

The standard deviation of adjacent correlations is:
\begin{equation}
\begin{split}
    \text{adjacent\_correlation\_std} = & \\
    \sqrt{
    \frac{1}{K-2}
    \sum_{k=1}^{K-1} 
    \left(\rho_k - \text{adjacent\_correlation\_mean}\right)^2
    }
\end{split}
\label{eq:adj-subc-std}
\end{equation}

\textbf{Spectral Roughness Analysis:} The first-order spectral difference (roughness) is defined as:
\begin{equation}
    \Delta |H(f_k)| = |H(f_{k+1})| - |H(f_k)|
    \label{eq:roughness}
\end{equation}
From this sequence, the mean and standard deviation of spectral roughness are computed as:

\begin{equation}
\begin{split}
    \text{spectral\_roughness\_mean} = 
    \frac{1}{K-1} \sum_{k=1}^{K-1} |\Delta |H(f_k)||
\end{split}
\end{equation}

\begin{equation}
\begin{split}
    \text{spectral\_roughness\_std} = & \\
    \sqrt{
    \frac{1}{K-2}
    \sum_{k=1}^{K-1}
    \left(|\Delta |H(f_k)|| - \text{spectral\_roughness\_mean}\right)^2
    }
\end{split}
\label{eq:spec-rough-std}
\end{equation}

\textbf{Spectral Curvature:} Spectral curvature is computed as the second-order finite difference:
\begin{equation}
    \Delta^2 |H(f_k)| = |H(f_{k+2})| - 2|H(f_{k+1})| + |H(f_k)|
    \label{eq:spec-curve}
\end{equation}
The mean and standard deviation of spectral curvature are then given by:

\begin{equation}
\begin{split}
    \text{spectral\_curvature\_mean} =
    \frac{1}{K-2} \sum_{k=1}^{K-2} |\Delta^2 |H(f_k)||
\end{split}
\end{equation}

\begin{equation}
\begin{split}
    \text{spectral\_curvature\_std} = & \\
    \sqrt{
    \frac{1}{K-3}
    \sum_{k=1}^{K-2}
    \left(|\Delta^2 |H(f_k)||
    - \text{spectral\_curvature\_mean}\right)^2
    }
\end{split}
\label{eq:spec-curve-std}
\end{equation}

\section{Beyond the results}
\label{app:beyond-results}

This appendix presents additional performance metrics and visualizations for the remaining classifiers evaluated in Section~\ref{sec:results}, serving to visually complement the comparative security analysis across models provided in the main text.

\vspace{10pt}

\noindent\textbf{Confusion Matrix Evaluation:}

\begin{figure}[htpb]
    \centering
    \begin{subfigure}[b]{0.48\columnwidth}
        \centering
        \includegraphics[width=\linewidth]{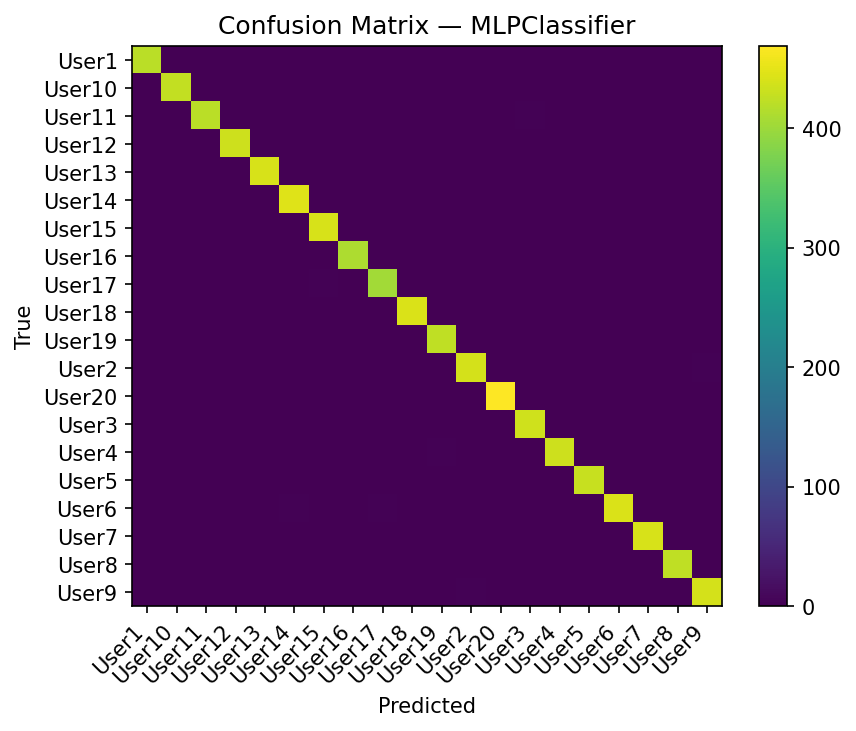}
        \caption{50-Sample Window Size}
        \label{fig:MLP-CM}
    \end{subfigure}
    \hfill 
    \begin{subfigure}[b]{0.48\columnwidth}
        \centering
        \includegraphics[width=\linewidth]{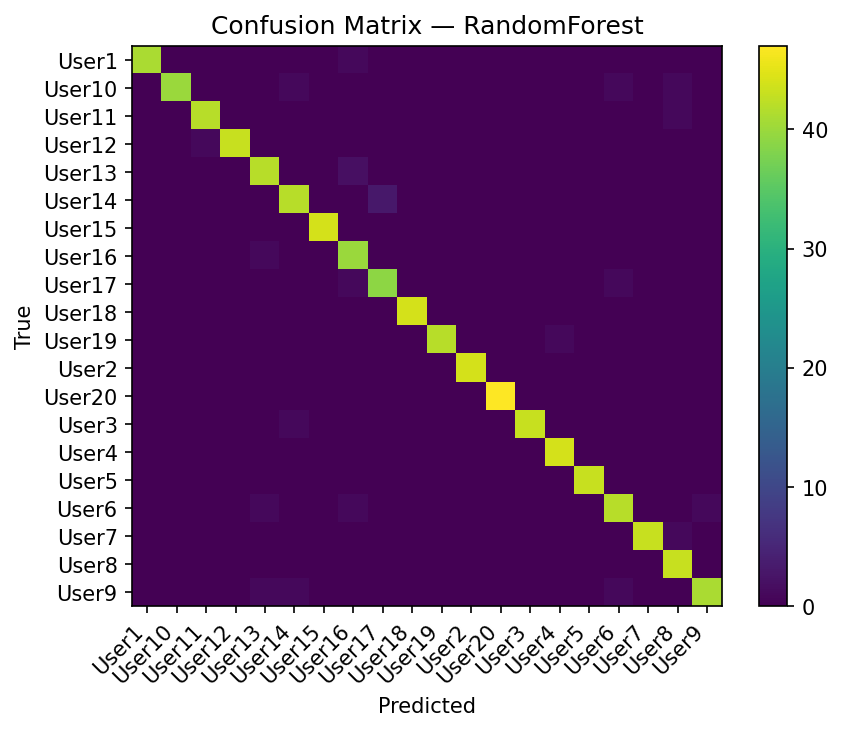}
        \caption{500-Sample Window Size}
        \label{fig:RF-CM}
    \end{subfigure}
    \caption{Confusion Matrix Comparison: MLP vs. Random Forest.}
    \label{fig:CM}
\end{figure}

\vspace{170pt}

\noindent\textbf{EER Metric: Random Forest and SVM (50-Sample Window):}

\begin{figure}[htbp]
    \centering
    \includegraphics[width=0.85\linewidth, keepaspectratio]{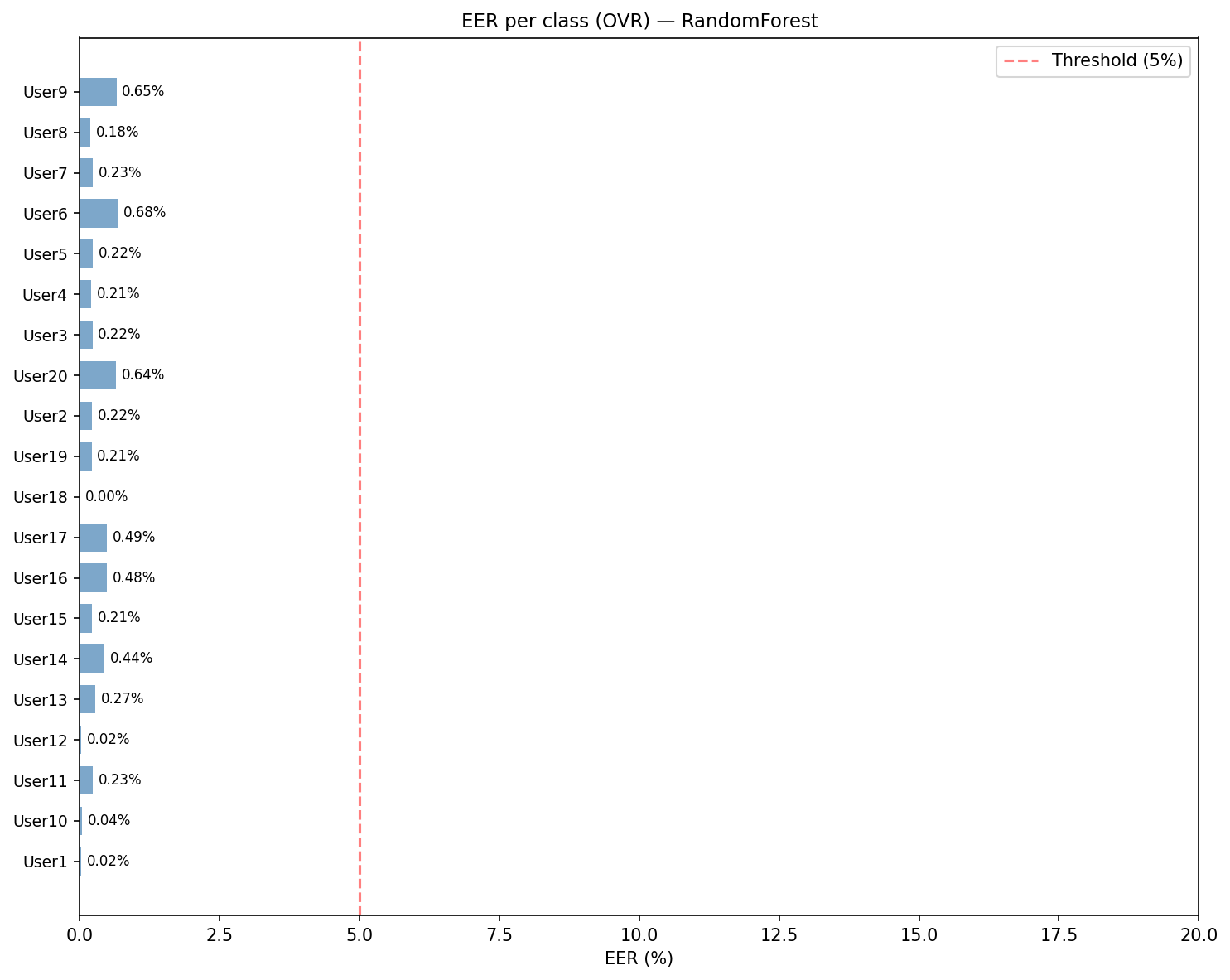}
    \caption{EER from Random Forest.}
    \label{fig:eer-RF-50}
\end{figure}

\begin{figure}[htbp]
    \centering
    \includegraphics[width=0.85\linewidth, keepaspectratio]{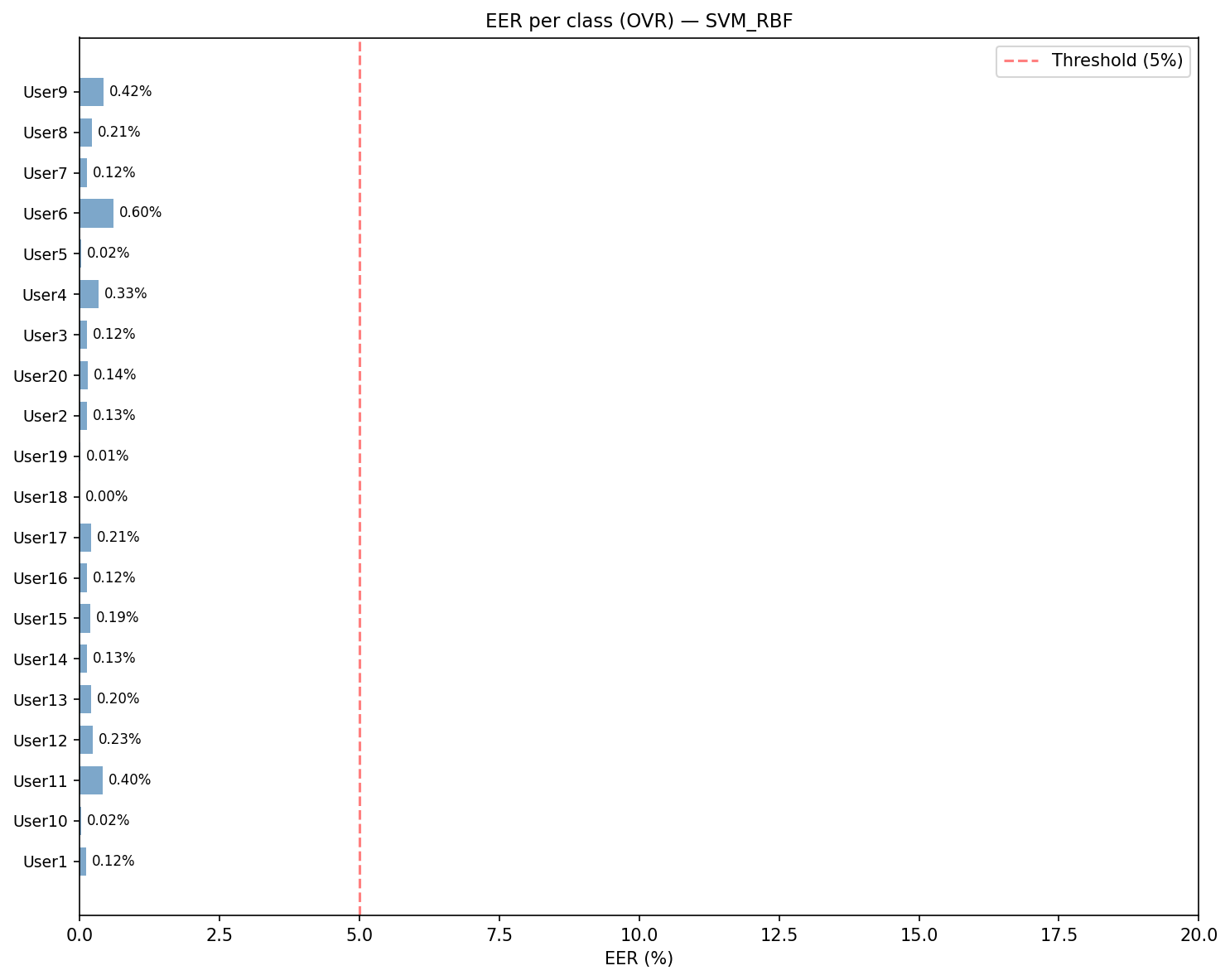}
    \caption{EER from SVM.}
    \label{fig:eer-SVM-50}
\end{figure}

\noindent\textbf{EER Metric: MLP, Random Forest and SVM (500-Sample Window):}

\begin{figure}[htbp]
    \centering
    \includegraphics[width=0.85\linewidth, keepaspectratio]{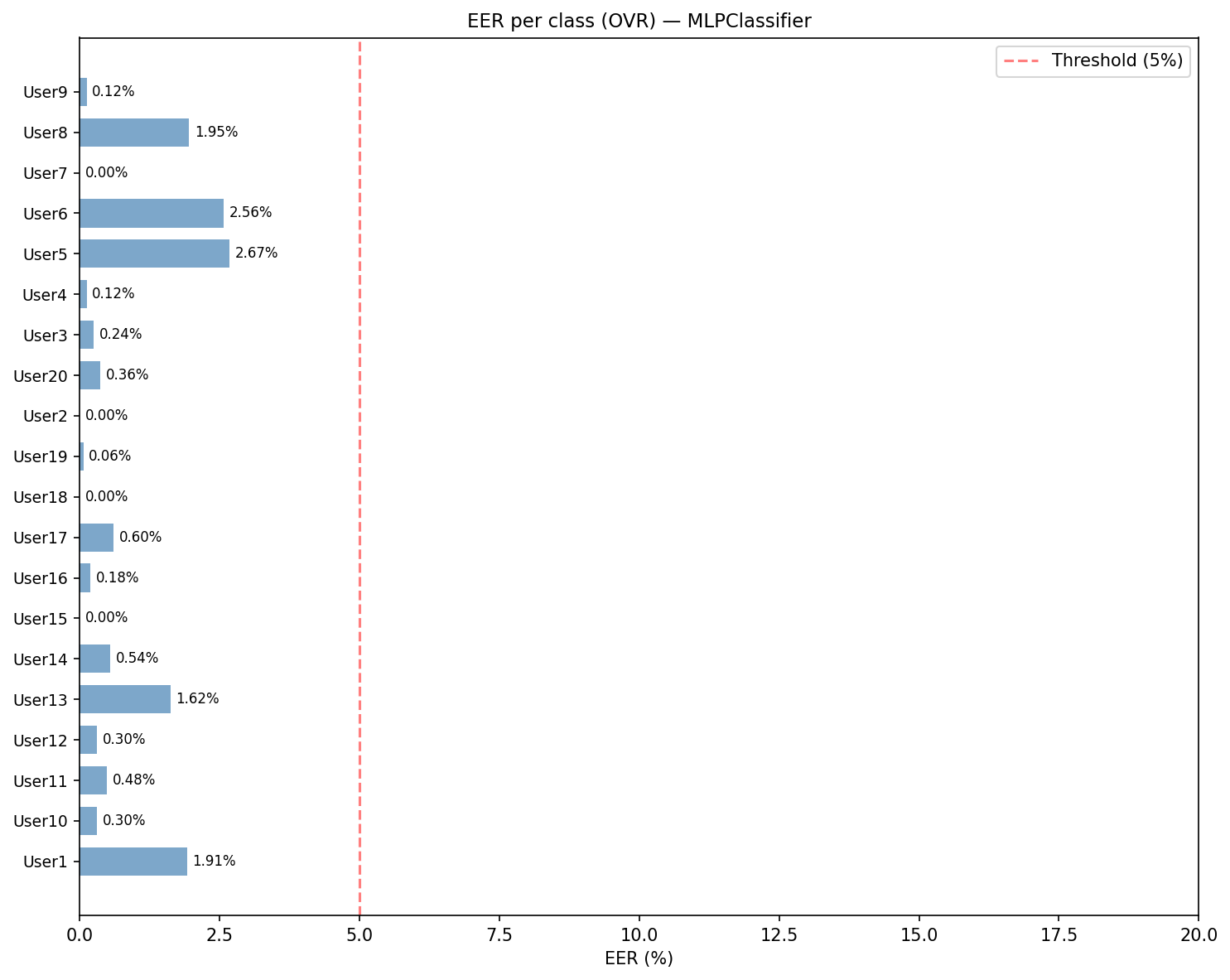}
    \caption{EER from MLP.}
    \label{fig:eer-MLP-500}
\end{figure}

\begin{figure}[ht]
    \centering
    \includegraphics[width=0.85\linewidth, keepaspectratio]{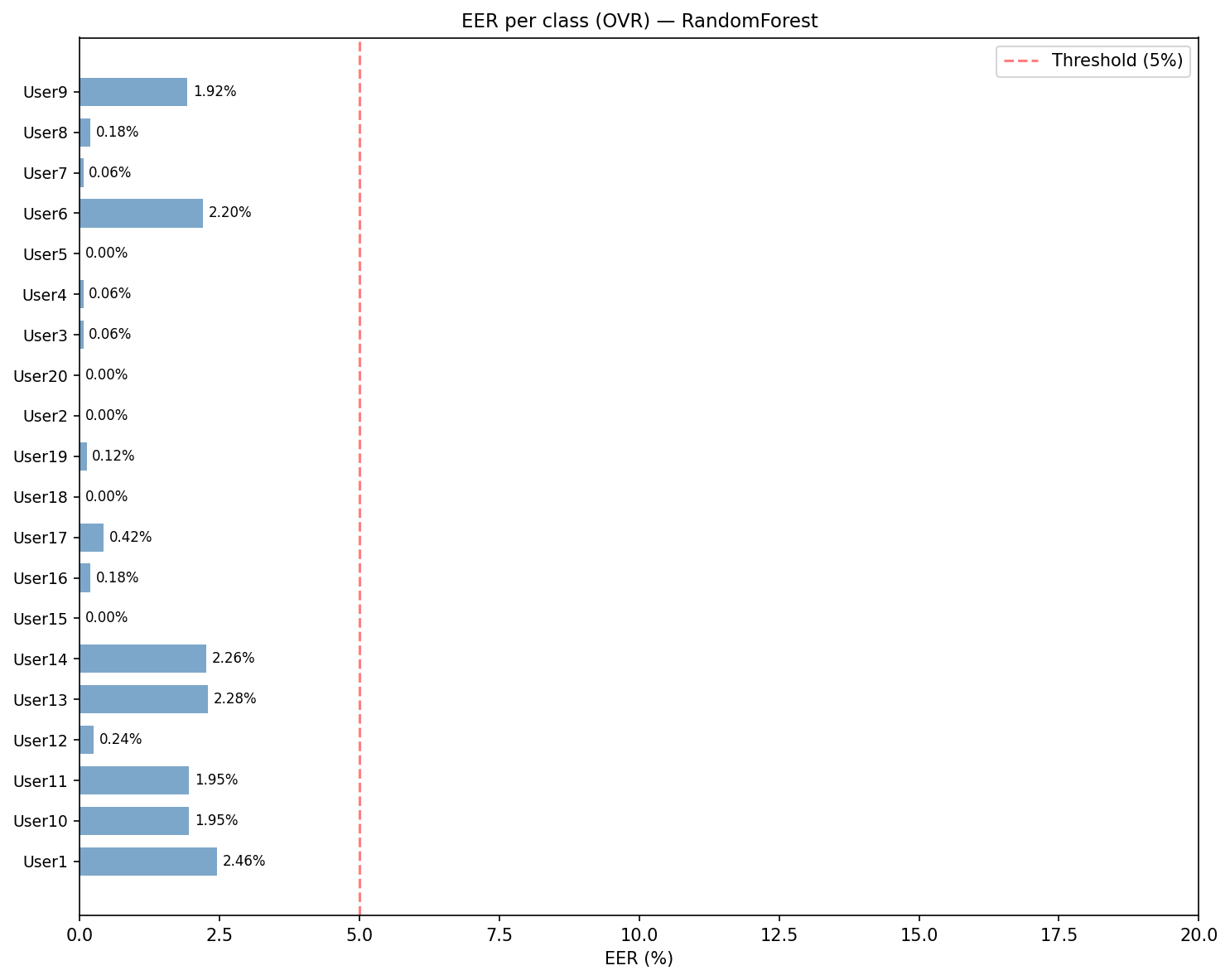}
    \caption{EER from Random Forest.}
    \label{fig:eer-RF-500}
\end{figure}

\begin{figure}[htbp]
    \centering
    \includegraphics[width=0.85\linewidth, keepaspectratio]{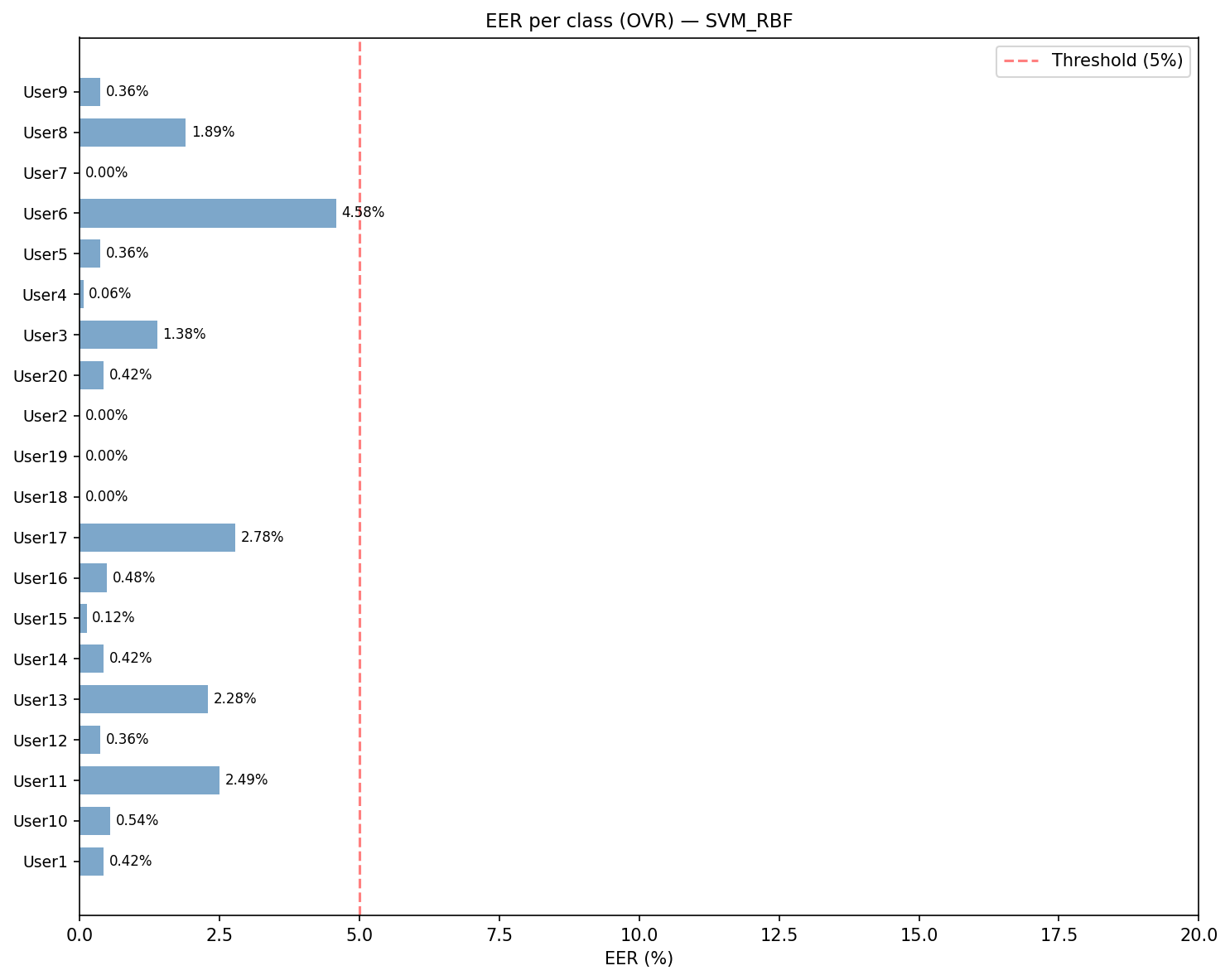}
    \caption{EER from SVM.}
    \label{fig:eer-SVM-500}
\end{figure}

\vspace{90pt}

\noindent\textbf{FCS Metric: Random Forest and SVM (50-Sample Window):}

\begin{figure}[htbp]
    \centering
    \includegraphics[width=0.9\linewidth, keepaspectratio]{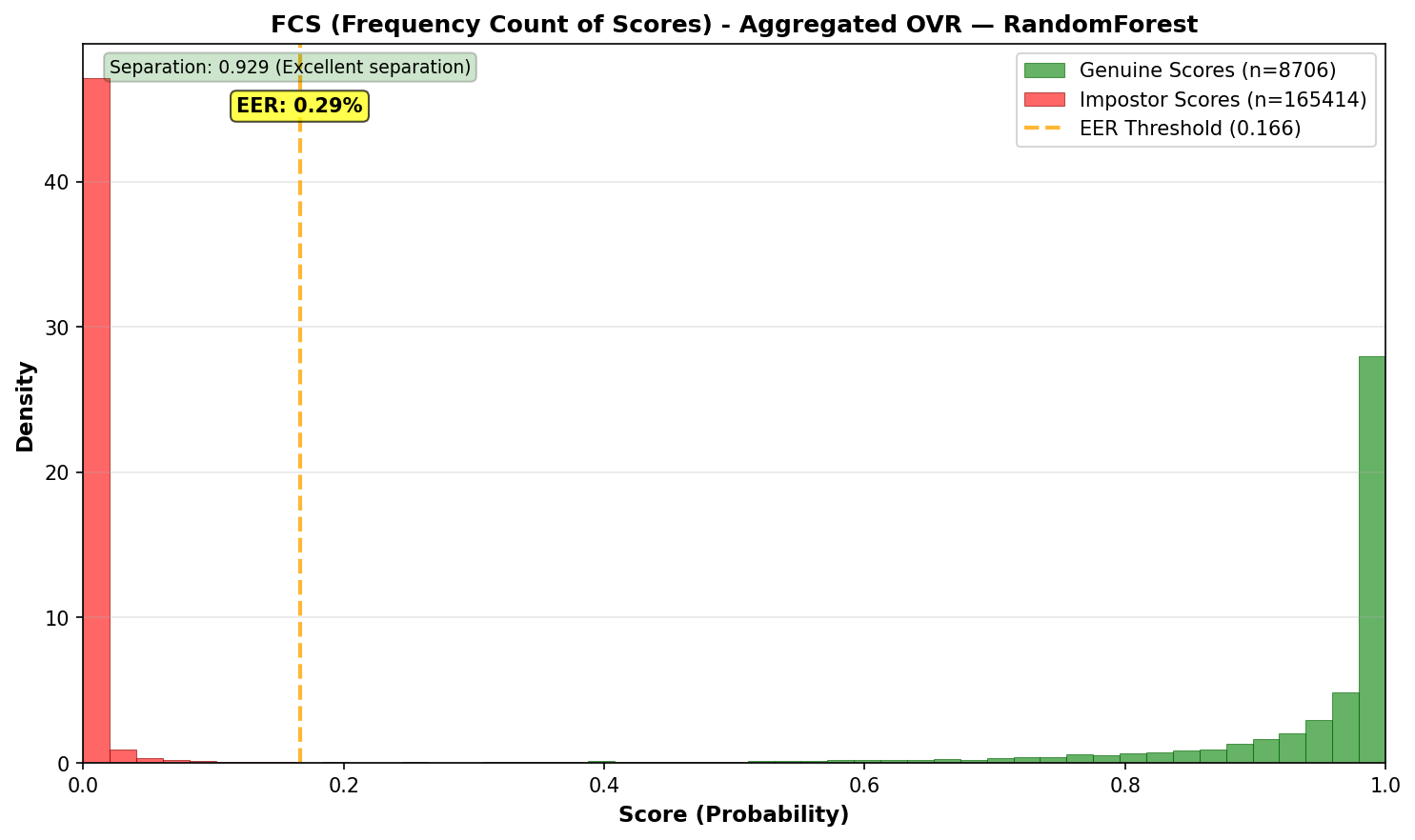}
    \caption{GC from Random Forest.}
    \label{fig:fcs-RF-500}
\end{figure}

\begin{figure}[htbp]
    \centering
    \includegraphics[width=0.9\linewidth, keepaspectratio]{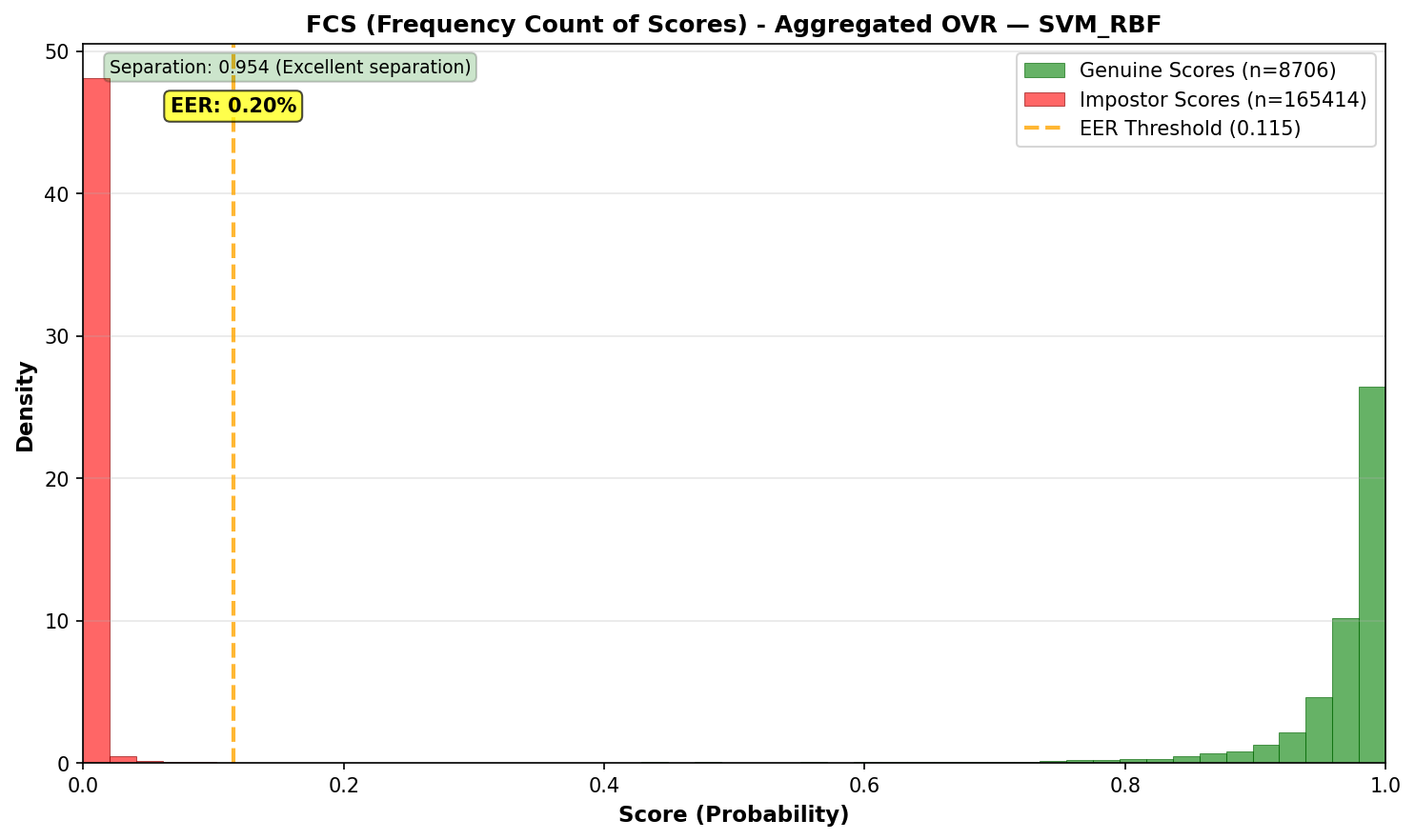}
    \caption{GC from SVM.}
    \label{fig:fcs-SVM-500}
\end{figure}

\vspace{160pt}

\noindent\textbf{Gini Metric: MLP, Random Forest and SVM (50-Sample Window):}

\begin{figure}[htbp]
    \centering
    \includegraphics[width=0.9\linewidth, keepaspectratio]{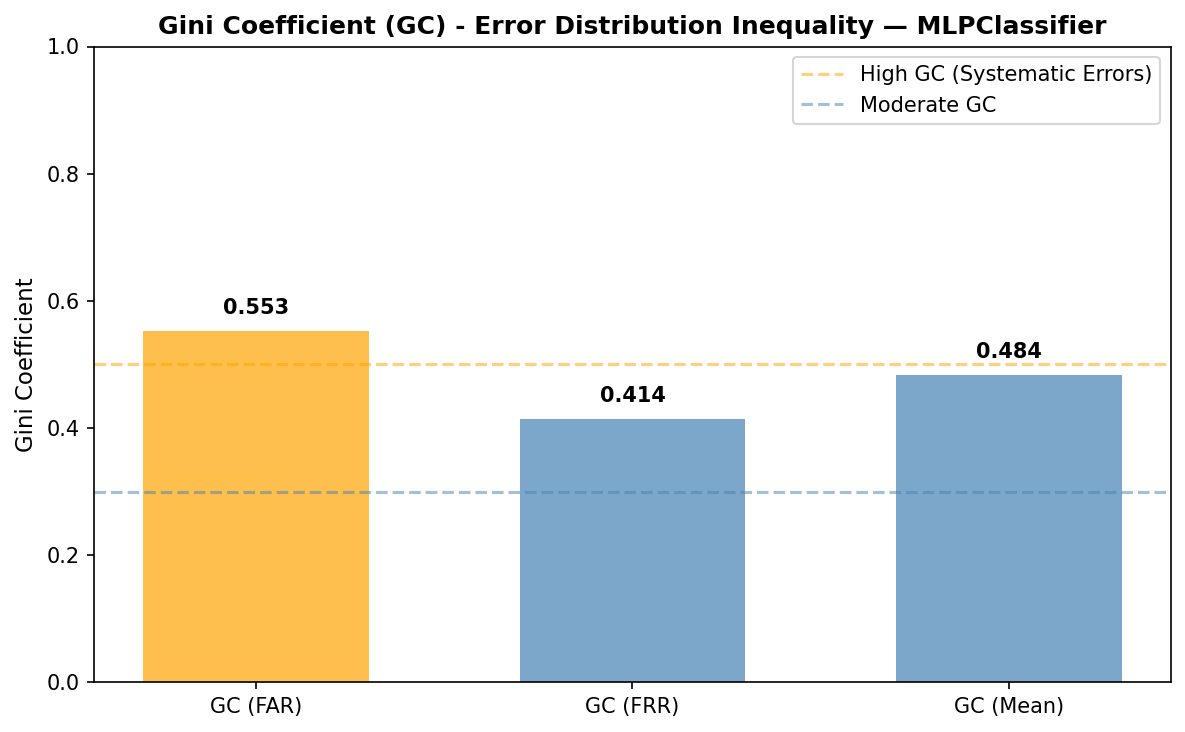}
    \caption{GC from MLP.}
    \label{fig:gc-MLP-500}
\end{figure}

\begin{figure}[htbp]
    \centering
    \includegraphics[width=0.9\linewidth, keepaspectratio]{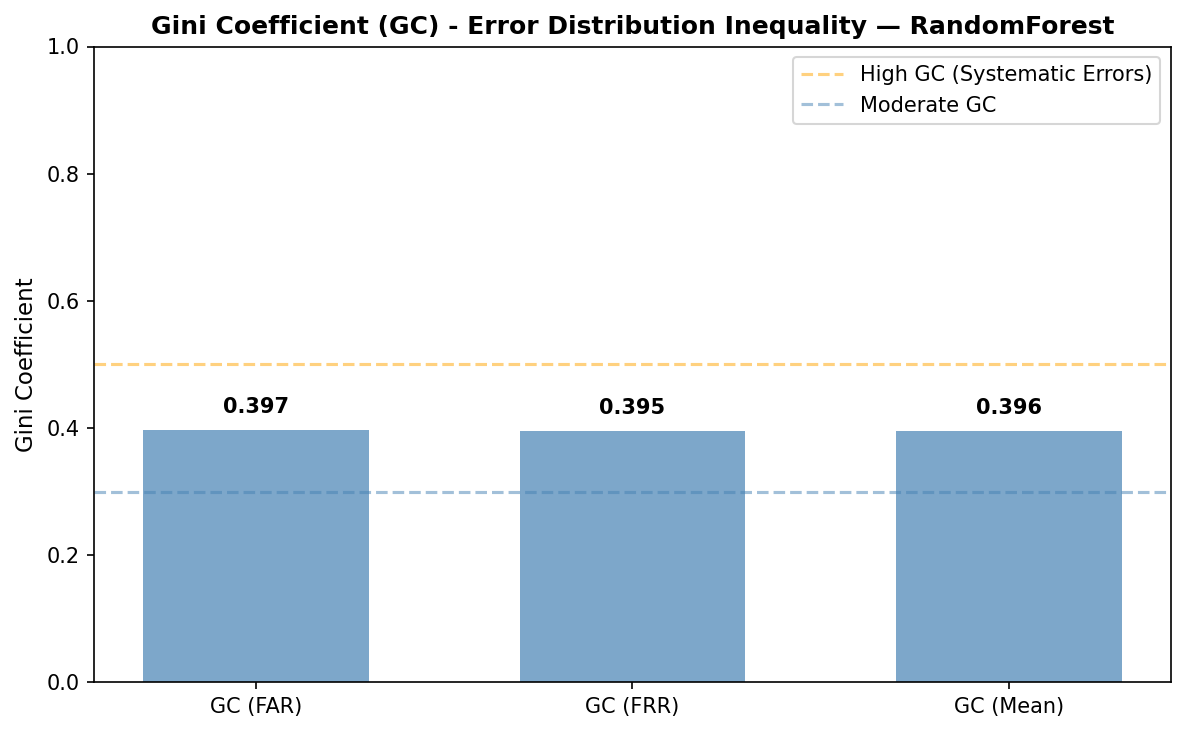}
    \caption{GC from Random Forest.}
    \label{fig:gc-RF-500}
\end{figure}

\begin{figure}[htbp]
    \centering
    \includegraphics[width=0.9\linewidth, keepaspectratio]{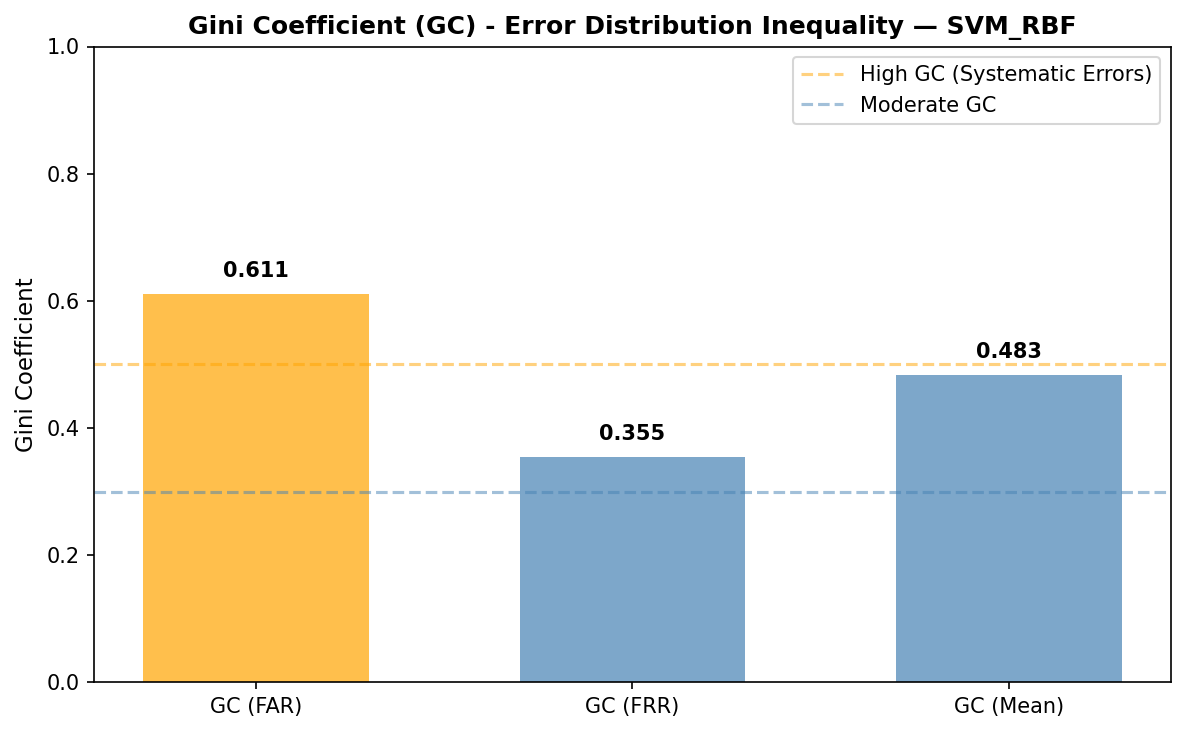}
    \caption{GC from SVM.}
    \label{fig:gc-SVM-500}
\end{figure}

\vspace{200pt}

\noindent\textbf{BioQuake Metric Evaluation Across Models (500-Sample Window):}

\begin{figure}[htbp]
    \centering
    \includegraphics[width=0.9\linewidth, keepaspectratio]{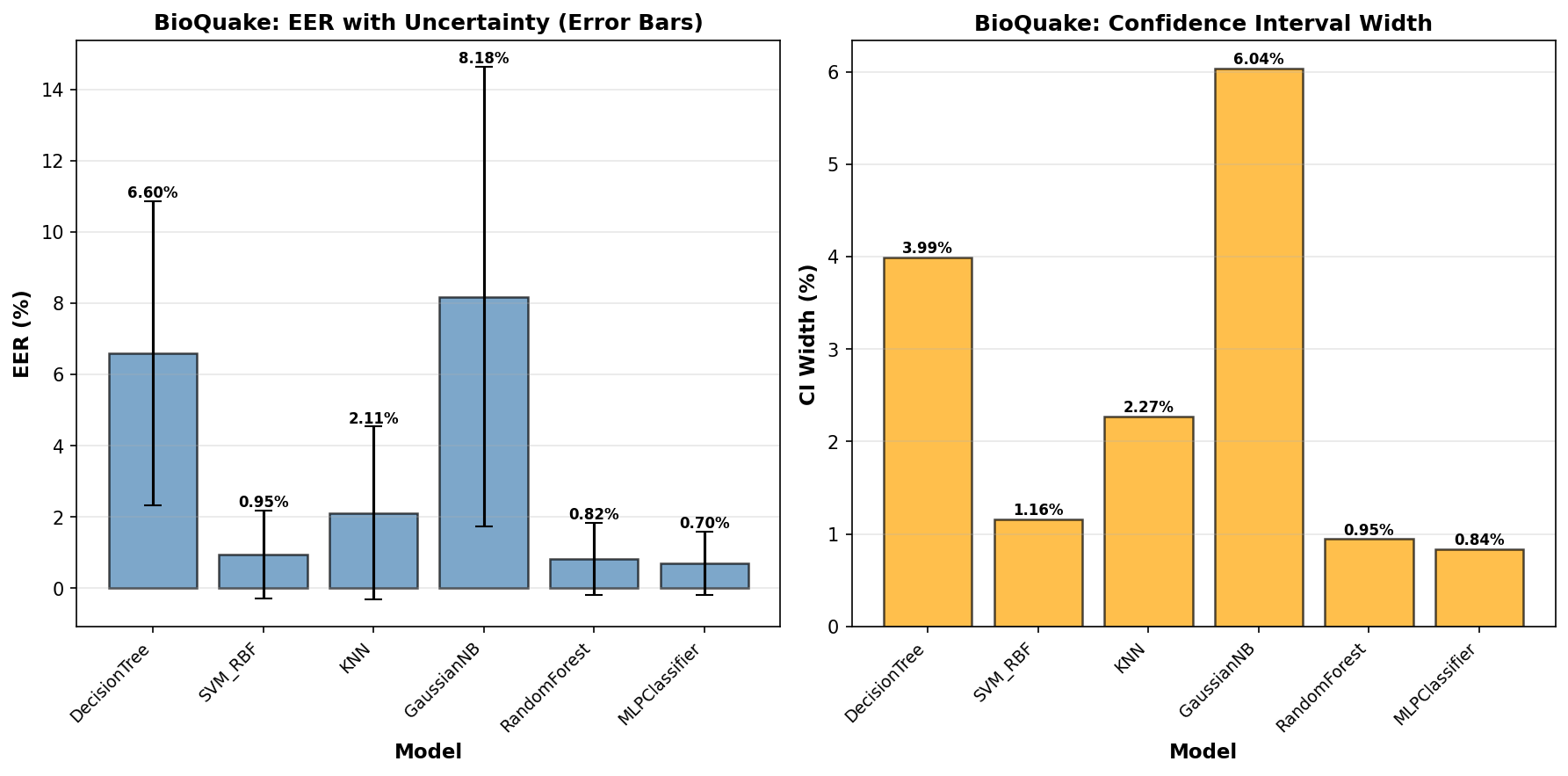}
    \caption{BioQuake Evaluation}
    \label{fig:gc-SVM-500}
\end{figure}

\end{document}

%% file: benchmark_table.tex
\begin{table*}[!h]

\centering
\caption{Benchmark Comparison of State-of-the-Art CSI-Based Biometric Authentication/Recognition Systems.\\ {\scriptsize \CIRCLE: \textbf{Yes} $|$ \LEFTcircle: \textbf{Partial} $|$ \Circle: \textbf{No}}}
\label{tab:benchmark_literature}
\renewcommand{\arraystretch}{0.8}
\setlength{\tabcolsep}{1pt}
\resizebox{\textwidth}{!}{%
\begin{tabular}{ccccccc c c c c c}
\toprule
\textbf{Work} & \textbf{Application} & \textbf{Dev. \& Tool} & \textbf{Freq.} & \textbf{Model} & \textbf{\makecell[c]{Passive\\Mode}} & \textbf{\makecell[c]{Static\\Acquisition}} & \textbf{\makecell[c]{Handcrafted\\Features}} & \textbf{\makecell[c]{EER\\FAR-FRR}} & \textbf{\makecell[c]{ROC-\\AUC}} & \textbf{\makecell[c]{Amp./\\Phase}} & \textbf{\makecell[c]{Best\\Perf. (\%)}} \\
\midrule
Lin et al. (2018) ~\cite{Lin2018} & \makecell[c]{Behavioral\\User Auth.} & \makecell[c]{TP-Link TL-WR886N \&\\Intel 5300 CSI \&\\Intel CSI} & 2.4~GHz & CNN + ResNet & \CIRCLE & \Circle & \Circle & \Circle & \Circle & Amp. & 98\% \\
\midrule
Kong et al. (2021)~\cite{Kong2021} & \makecell[c]{Behavioral\\Multi-User Auth.} & \makecell[c]{Atheros NIC \&\\Atheros CSI} & 2.4 \& 5~GHz & CNN + RNN & \Circle &  \Circle & \LEFTcircle & \CIRCLE & \Circle & Both & 87.6\% \\
\midrule
Shi et al. (2021)~\cite{Shi2021} & \makecell[c]{Phys. \& Behav.\\User Auth.} & \makecell[c]{Intel 5300 NIC \&\\Intel 5300 CSI} & 2.4~GHz & \makecell[c]{Deep Learning +\\Transfer Learning} & \CIRCLE &  \Circle & \LEFTcircle &  \Circle & \Circle & Both & 96.8\% \\
\midrule
Gu et al. (2021)~\cite{Gu2021} & \makecell[c]{Behavioral\\User Auth.} & \makecell[c]{Intel 5300 NIC \&\\Linux CSI} & 2.4 \& 5~GHz & Deep Learning & \CIRCLE &  \Circle & \LEFTcircle  & \CIRCLE & \Circle & Amp. & 94.3\% \\
\midrule
Gu et al. (2022)~\cite{Gu2022} & \makecell[c]{Behavioral\\ User Auth.} & \makecell[c]{Intel 5300 NIC \&\\Intel CSI Tool} & 2.4~GHz & CNN + SVM & \CIRCLE &  \Circle & \LEFTcircle  & \CIRCLE & \Circle & Amp. & 92.1\% \\
\midrule
Wang et al. (2022)~\cite{Wang2022} & \makecell[c]{Gait \& Respiration\\Recognition} & \makecell[c]{Intel 5300 NIC \&\\Intel 5300 CSI} & 2.4~GHz & Deep Learning & \Circle &  \Circle & \LEFTcircle & \CIRCLE & \Circle & Amp. & 98.75\% \\
\midrule
Turetta et al. (2022)~\cite{Turetta2022} & \makecell[c]{Behavioral\\User Identif.} & \makecell[c]{RPi 4B \&\\Nexmon} & 2.4~GHz & CNN & \CIRCLE &  \Circle & \Circle & \Circle & \Circle & Amp. & 98.2\% \\
\midrule
Zhang et al. (2022)~\cite{Zhang2022} & \makecell[c]{Gesture\\Recognition} & \makecell[c]{Intel 5300 NIC \&\\Linux CSI} & 5~Ghz & CNN + GRU & \CIRCLE &  \Circle & \Circle & \Circle & \Circle & Both & 92.7\% \\
\midrule
Geng et al. (2022)~\cite{Geng2022} & \makecell[c]{Pose\\Estimation} & \makecell[c]{TP-Link AC1750 \&\\CSI Tool} & 2.4~GHz & \makecell[c]{Deep Learning +\\RCNN + ResNet} & \CIRCLE &  \Circle & \Circle & \Circle & \Circle & Both & 87.2\% \\
\midrule
Lin et al. (2023)~\cite{Lin2023} & \makecell[c]{Phys. \& Behav.\\User Auth.} & \makecell[c]{TP-Link TL-WR886N \&\\Intel 5300 CSI \&\\Intel CSI} & 2.4~GHz & \makecell[c]{Deep Learning +\\Transfer Learning} & \CIRCLE &  \Circle & \LEFTcircle & \Circle & \CIRCLE & Both & 97.1\% \\
\midrule
Kong et al. (2023)~\cite{Kong2023} & \makecell[c]{Beravioral\\User Auth.} & \makecell[c]{Intel 5300 \&\\Atheros CSI} & 5~GHz & CNN-RNN + GRU & \CIRCLE &  \Circle & \LEFTcircle & \CIRCLE & \Circle & Both & 86.2\% \\
\midrule
Guan et al. (2023)~\cite{Guan2023} & \makecell[c]{Human Occupancy\\Counting} & \makecell[c]{RPi 4B \&\\Nexmon} & 2.4 \& 5~GHz & SVM & \Circle &  \Circle & \LEFTcircle & \Circle & \CIRCLE & Amp. & 96.0\% \\
\midrule
Bisio et al. (2024)~\cite{Bisio2024} & \makecell[c]{Human Act. Recog.\\HAR} & \makecell[c]{Intel 5300 NIC \&\\Intel 5300 CSI} & 5~GHz & Random Forest & \Circle &  \Circle & \LEFTcircle & \CIRCLE & \Circle & Both & 99.92\% \\
\midrule
Li et al. (2024)~\cite{Li2024.2} & \makecell[c]{Gesture\\Recognition} & \makecell[c]{Intel AX210 NIC\\Linux (debugfs) + MatLab} & 2.4--7.1 GHz & \makecell[c]{DenseNet + SKNet +\\Transformer attention} & \Circle &  \Circle & \Circle & \Circle & \Circle & Both & 98.6\% \\
\midrule
Trindade et al. (2025)~\cite{trindade2025} & \makecell[c]{Phys. \& Biomet.\\User Auth.} & \makecell[c]{TP-Link Archer C60 \&\\RPi 4B \&\\Nexmon} & 5~GHz & Random Forest & \CIRCLE &  \CIRCLE & \Circle & \Circle & \Circle & Both & 99.81\% \\
\midrule
Ramesh et al. (2025)~\cite{Ramesh2025} & \makecell[c]{Image\\Reconstruction} & \makecell[c]{Intel AX210 Wi-Fi 6 \&\\FeitCSI tool \&\\Cam. Intel Realsense D435} & 5~GHz & \makecell[c]{Deep Learning +\\Diffusion Models} & \Circle &  \CIRCLE & \Circle & \Circle & \Circle & Amp. & 92.0\% \\
\bottomrule
\end{tabular}
}
\end{table*}